\documentclass[a4paper]{article}

\usepackage[margin=1in]{geometry} % full-width

\usepackage{amsmath}
\usepackage{comment}
\usepackage{amsthm}
\usepackage{amssymb}
\usepackage[table,xcdraw]{xcolor}
\usepackage{longtable}
\usepackage{float}
\usepackage{wasysym}
\usepackage[utf8]{inputenc}
\usepackage{hyperref}
\usepackage{oplotsymbl}
\usepackage{graphicx, color}
\usepackage{enumitem}

\usepackage{mathrsfs} % for \mathscr command

\usepackage{lipsum}
\usepackage[T1]{fontenc}
\usepackage{comment}
\usepackage{multirow}
\usepackage{longtable}
\usepackage{lscape}
\usepackage{url}
\usepackage{subcaption}
\usepackage{booktabs}
\usepackage[table,xcdraw]{xcolor}
\usepackage{longtable}

\usepackage{bbm}
\usepackage{bm}

\usepackage[numbers]{natbib}
\title{Cryptocurrencies in the Balance Sheet: Insights from (Micro)Strategy - Bitcoin Interactions
}

\author{\small Sabrina Aufiero$^{1, *}$, Antonio Briola$^{1}$, Tesfaye Salarin$^{2}$, Silvia Bartolucci$^{1}$, Fabio Caccioli$^{1,3}$, Tomaso Aste$^{1}$ \\  }

\date{ \small $^1$ Department of Computer Science, University College London, 66-72 Gower Street, WC1E 6EA, London, UK \\ $^2$ Department of Economics, University of Trento, 5 Vigilio Inama, 38122, Trento, IT \\ $^3$ Systemic Risk Centre, London School of Economics and Political Sciences, Houghton Street, WC2A 2AE, London, UK \\ $^\star$ Corresponding author: \url{sabrina.aufiero.22@ucl.ac.uk}
}

\begin{document}

\flushbottom
\maketitle
%\tableofcontents

\begin{abstract}
This paper investigates the evolving link between cryptocurrency and equity markets in the context of the recent wave of corporate Bitcoin (BTC) treasury strategies. We assemble a dataset of $39$ publicly listed firms holding BTC, from their first acquisition through April $2025$. Using daily logarithmic returns, we first document significant positive co-movements via Pearson correlations and single factor model regressions, discovering an average BTC beta of $0.62$, and isolating $12$ companies, including \textit{Strategy} (formerly \textit{MicroStrategy}, MSTR), exhibiting a beta exceeding $1$. We then classify firms into three groups reflecting their exposure to BTC, liquidity, and return co-movements. We use transfer entropy (TE) to capture the direction of information flow over time. Transfer entropy analysis consistently identifies BTC as the dominant information driver, with brief, announcement-driven feedback from stocks to BTC during major financial events. Our results highlight the critical need for dynamic hedging ratios that adapt to shifting information flows. These findings provide important insights for investors and managers regarding risk management and portfolio diversification in a period of growing integration of digital assets into corporate treasuries.
\end{abstract}

\section{Introduction}\label{sec:Introduction}

Bitcoin (BTC-USD pair, hereafter BTC) emerged as the top-performing asset in $2024$, a phenomenon largely attributable to the launch of spot exchange-traded funds (ETFs) and growing optimism regarding potential regulatory easing under the newly inaugurated U.S. administration. Over the course of $2024$, BTC delivered a return exceeding $113\%$, significantly outperforming all major traditional asset classes. In comparison, the S\&P 500 (SPY) returned $23.7\%$, gold (GLD) $28.7\%$, government bonds (GOVT) $-2.18\%$, and real estate (VNQ) $-0.93\%$ \cite{forbes2025bitcoin2024}.

Despite this remarkable performance, BTC remains a highly volatile asset. By the end of $2024$, its price had more than doubled from approximately $\$40,000$ to nearly $\$94,000$. The most pronounced acceleration occurred following the U.S. presidential election, with BTC surpassing the $\$108,000$ threshold by mid-December. This surge was widely interpreted as a market response to expectations that President Donald Trump’s victory would have increased institutional participation in the cryptocurrency market.

In addition to these exogenous factors, a key structural feature influencing BTC price dynamics is the halving mechanism embedded in the BTC protocol \cite{meynkhard2019fair, m2024bubbles}. Approximately every four years, or after every $210,000$ blocks, the reward granted to miners for validating transactions and appending new blocks is reduced by half. The primary purpose of this mechanism is to regulate the supply of new BTCs, ultimately capping the total number of coins at $21$ million. Historical evidence shows the economic significance of these events: following the first halving on November $28$, $2012$, when the block reward decreased from $50$ to $25$ BTC, a substantial price escalation ensued. The most recent halving occurred on April $20$, $2024$, at block $740,000$, reducing the reward from $6.25$ BTC to $3.125$ BTC. Empirical studies suggest that such supply contractions contribute to increased scarcity, reinforcing BTC’s role as a digital store of value -- frequently compared to gold -- and fostering investors' interest and trading activity \cite{yahoofinance2025mara}.

Broader trends in global investment strategies demonstrate the increasing integration of BTC into traditional financial markets. By the end of $2024$, the value of globally invested assets surpassed $\$200$ trillion, with cryptocurrencies representing $1.5\%$ of the total, with approximately $\$3$ trillion \cite{ReutersCrypto2024}. Institutional investors substantially increased their exposure to spot BTC ETFs in the fourth quarter of $2024$, with reported holdings reaching $\$38.7$ billion -- more than three times the $\$12.4$ billion recorded in the previous quarter \cite{YahooFinanceBitcoinETFs2025}.
This growing institutional involvement has reinforced the dual role of BTC as both a benchmark for the broader cryptocurrency market, and an indicator of the global risk appetite. Its finite supply and decentralized governance contrast sharply with traditional financial instruments, positioning BTC as an alternative growth asset within a changing economic environment. Discussions regarding the potential classification of BTC as a strategic reserve asset in the U.S. -- in a period of rising public debt and increased borrowing costs -- further underline the cryptocurrency's evolving role in the global financial system. 

An episode happened in April $2025$ on Wall Street highlights how firmly digital assets have integrated into mainstream financial markets. Following the circulation of a false headline suggesting that the U.S. President Donald Trump would have delayed planned tariffs by $90$ days, the S\&P 500 temporarily gained approximately $\$2.5$ trillion in market value, only to reverse those gains once the information was disproved. BTC and other major cryptocurrencies mirrored this sharp rise and subsequent fall, reacting to macroeconomic rumors and exhibiting the same risk-on/risk-off behavior traditionally associated with equity markets \cite{lipschultz2025madness}. Such dynamics confirm that cryptocurrencies now participate in the global feedback loops that govern conventional financial assets, marking their transition from speculative instruments to a fully fledged asset class. 

During the Federal Reserve’s 2022--2024 tightening cycle, BTC increasingly behaved like other speculative growth assets, particularly the so-called high-beta technology stocks \cite{mijares2024bitcoin}. Recent data support this view. The 90-day rolling correlation between BTC and the NASDAQ-$100$ index reached $0.46$ in May $2024$ -- the highest since August $2023$ -- and had previously spiked above $0.80$ in early $2022$ after the Fed’s initial rate hikes \cite{mijares2024bitcoin}. These trends challenge the traditional view of BTC as a store of value immune to macroeconomic swings. Its sharp price fluctuations, combined with the launch and rapid growth of U.S. spot BTC ETFs, have broadened its investor base and further linked its returns to overall market risk sentiment. BTC now behaves more as a traditional growth asset, reflecting its evolving role in institutional portfolios.

Parallel to the increasing institutional adoption of BTC, a small but growing set of listed firms have begun to re-orient their corporate treasuries around direct cryptocurrency holdings. The most prominent case is \textit{Strategy Incorporated} (formerly \textit{MicroStrategy Incorporated}, hereafter MSTR), that offers AI powered enterprise analytics software and services. Its Executive Chairman Michael Saylor has articulated a business model that treats BTC as the firm’s primary reserve asset. Since August $2020$ \textit{Strategy} has financed successive purchases of BTC through a mix of follow-on equity offerings, senior secured notes, and convertible bonds, supplementing residual cash flows from its legacy software division \cite{phillips2024microstrategy}. As on April $27$, $2025$, the company held $553,555$ BTC, the most prominent corporate position worldwide, and Mr. Saylor ranked as the top percentage gainer on Forbes’ $2024$ list of crypto billionaires \cite{ForbesCryptoBillionaires2024}. Although the underlying software business has shown only modest growth, the equity has attracted substantial retail and institutional demand precisely because of the balance sheet's aggressive exposure to BTC \cite{pan2025saylor}.

Imitative strategies have emerged in other jurisdictions. \textit{Metaplanet Inc.}, often described as the ``Japanese \textit{Strategy}'', sold most of its hospitality real estate assets in $2024$ and redirected the proceeds, together with borrowed funds, into BTC; its flagship property in Tokyo has been re-branded ``The Bitcoin Hotel''. Over the twelve months to April $2025$, \textit{Metaplanet}’s share price rose roughly $4,800\%$, the most significant gain among Japanese equities and one of the highest globally \cite{bloomberg2025metaplanet}. 
In the United States, \textit{Fold Holdings Inc.} is the first publicly traded financial services company built entirely around BTC, offering rewards, savings, and payment solutions denominated in BTC. On July $24$, $2024$, \textit{Fold} executed its inaugural BTC purchase of $1,000$ BTC \cite{BitcoinTreasuries}.  Thereafter, the company issued a $\$46.3$ million convertible note secured by $500$ BTC and added $475$ BTC and a further $10$ BTC to its treasury, placing it among the top ten U.S. public firms by crypto holdings.  Despite this aggressive accumulation strategy, \textit{Fold}’s share price has fallen by $62.3\%$ over the past year \cite{investing2025fld}.

These and similar convertibles have become attractive to hedge funds seeking volatility-driven arbitrage. Large investors restricted from holding BTC directly likewise use such equities as indirect proxies \cite{pan2025saylor}. Smaller issuers have adopted analogous tactics with other tokens: \textit{Upexi Inc.} raised $\$100$ million in April $2025$ to acquire Solana (SOL), lifting its share price by more than $325\%$ on the announcement date, while \textit{Janover Inc.} (a real estate finance company controlled by a consortium of former Kraken executives) disclosed a $\$10.5$ million purchase of SOL earlier the same month \cite{ghosh2025solana}.

Despite their appeal, these ``crypto treasury'' models entail significant financial risk. Digital asset price volatility can translate into earnings variability and elevated refinancing risk, the effects of which are amplified when acquisitions are debt-financed. These concerns are especially acute for smaller firms, whose core operating businesses are only tangentially related to cryptocurrency markets, raising questions about their long-run solvency \cite{ghosh2025solana}. These examples highlight not only the growing corporate appetite for direct crypto exposure, but also the critical importance of robust risk management and disclosure practices when adopting such balance sheet strategies.

This study therefore investigates the evolving relationship between Bitcoin (BTC) and equity markets, focusing on firms that have adopted BTC as part of their corporate treasury strategy. Using a dataset of 39 publicly listed BTC-holding companies from $2017$ to $2025$, the paper applies correlation analysis, single factor return models, and transfer entropy methods to quantify both linear and nonlinear dependencies. We will show that BTC consistently acts as the dominant information driver, particularly during major market events, while feedback from equities to BTC remains rare and event-specific. By illustrating these directional and time-varying patterns, the study provides investors, risk managers, and policymakers useful information for managing the growing integration of digital assets into conventional financial markets.

The remainder of the paper is organized as follows. Section \ref{sec: Lit Review} reviews the literature on entropy based techniques in financial research. Section \ref{sec:Data} describes the dataset and provides a detailed analysis of MSTR, the specific stock of interest. Section \ref{sec:Methodology} outlines the empirical methodology, while Section \ref{sec:Results} presents the results from Pearson correlations, single factor model regressions, and transfer entropy analyses. Finally, Section \ref{sec:Conclusions} concludes the paper and discusses remaining open challenges.

\section{Literature Review}\label{sec: Lit Review}

Early empirical research on market interconnectedness has heavily relied on the correlation function to quantify co-movements among asset returns \cite{mantegna1999hierarchical, onnela2003dynamics, kullmann2002time, plerou1999universal, jung2006characteristics, briola2023anatomy, vidal2023ftx, briola2022dependency}. While convenient, correlation analysis suffers from two well-known shortcomings \cite{briola2024deep}. First, it captures only linear associations, thereby overlooking the nonlinear dependencies characterizing financial data. Second, correlation is intrinsically symmetric and, therefore, incapable of indicating which series, if any, exerts influence over the other.
Mutual Information (MI) provides an entropy-based measure of dependence sensitive to nonlinear structures, but, like correlation, remains directionless \cite{shannon1948mathematical}. Directionality can be recovered via Transfer Entropy (TE) \cite{bossomaier2016transfer, Aste_2025}, an extension of Wiener-Granger causality \cite{granger1969investigating, bressler2011wiener, lin2024granger} that quantifies the incremental information flow from one time series to another \cite{schreiber2000measuring}.

Marschinski and Kantz (2002) \cite{marschinski2002analysing} were among the first to apply TE in a financial context. They quantified the information flow between the Dow Jones and DAX index, identifying a clear unidirectional transfer from the U.S. market to the German one. Building on this foundation, Baek et al. (2005) \cite{baek2005transfer} employ TE to map pairwise information flows within the U.S. equity market, identifying energy sector firms -- particularly those engaged in oil, gas, and electricity -- as net transmitters of information to the broader market. Their results suggest that TE not only captures nonlinear dependencies but also helps to isolate market-leading institutions. Dimpfl et al. (2013) \cite{dimpfl2013using} use TE firstly to examine the importance of the credit default swap market relative to the corporate bond market for the pricing of credit risk, and secondly to analyze the dynamic relation between market risk and credit risk. Sandoval Jr (2014) \cite{sandoval2014structure} analyses daily returns for the $197$ largest global financial companies from $2003$ to $2012$. One-day lagged TE is used to map causal links, revealing which firms dominate others within the financial sub-sectors -- banks, diversified financial services, savings and loans, insurance, private equity funds, real estate investment companies, and real estate trust funds. He et al. (2017) \cite{he2017comparison} analyze the relationships among $9$ stock indices from the U.S., Europe, and China over the period $1995$--$2015$ using various forms of TE. They find that the U.S. holds the leading position in long term lagged relationships, whereas China emerges as the most influential market at shorter lags. Dimpfl et al. (2019) \cite{dimpfl2019group} introduce Effective Group transfer entropy (EGTE) as a tool for identifying informational leadership. They devise a bootstrap method for confidence intervals, demonstrate that linear techniques miss many information flows, and apply EGTE to intraday cryptocurrency data, uncovering predominantly nonlinear dependencies. Peng et al. (2022) \cite{peng2022pearson} compare simple Pearson correlations with TE for Chinese stocks using rolling windows and lead-lag shifts. Neto et al. (2023) \cite{neto2023ranking} use the same information measure to build a directed, weighted network of Brazilian equities on a $32$ year period in which each edge quantifies information flow from one stock’s returns to another. Applying network centrality measures, the authors successfully identify the most influential and the most influenced stocks at each point in time. Mungo et al. (2024) \cite{mungo2024cryptocurrency} demonstrate that investment structures themselves can shape market dynamics, as correlations in token returns tend to reflect the underlying patterns of crypto co-investment across investors. 

Ma (2025) \cite{ma2025volatility} examines the volatility dynamics of BTC and MSTR from September $2019$ to September $2024$ using the GARCH model \cite{bollerslev1986generalized}. They prove that MSTR's volatility has closely followed the footsteps of BTC, especially during the $2024$ rally in that market, including all its shocks and recoveries. Krause (2025) \cite{krause2025ponzi} finds that, while \textit{Strategy} has generated substantial returns, its debt- and equity-financed accumulation of BTC raises concerns about sustainability and speculative risks. Because the business model depends on continued capital inflows and a rising BTC price, several commentators liken it to a Ponzi-style structure. Krause evaluates the risk return profile of this strategy, benchmarks the firm’s performance against BTC itself, and weighs the relative merits of holding MSTR shares versus owning the underlying cryptocurrency directly.

\section{Data}\label{sec:Data}

Although no official registry exists, the public database ``Bitcoin Treasuries'' \cite{BitcoinTreasuries} reports that $66$ publicly listed companies worldwide -- and an additional $12$ private firms -- hold BTC on their balance sheets, including many whose core businesses were previously unrelated to cryptocurrencies. In this paper, we construct a dataset of BTC holders, including corporate and government entities that hold this crypto asset directly or through custodial arrangements, based on data from \cite{BitcoinTreasuries}, updated as of April $11$, $2025$. Our analysis focuses on $39$ publicly listed companies\,\footnote{For convenience, we use the terms `companies', `equities', and `stocks' interchangeably throughout the paper, although we primarily refer to the publicly traded shares of these companies.} that meet at least one of the following criteria: (i) they hold more than $700$ BTC, or (ii) they have a market capitalization exceeding $\$1.00B$, or (iii) their BTC holdings represent more than $50.00\%$ of that market capitalization. These thresholds ensure that each firm’s BTC exposure is economically significant and will likely have a measurable influence on its share price dynamics. The complete list of companies is presented in Table \ref{tab:tab dataset}, Appendix \ref{app:dataset}.

For each qualifying holder, we collect their BTC acquisition balance sheet data and match it with the firm’s market capitalization on the corresponding dates, as reported by the LSEG Refinitiv Eikon platform. We also retrieve daily closing share prices from the date of each firm’s initial BTC purchase through April $1$, $2025$\,\footnote{Daily closing prices are obtained with the Python package `yfinance'.}. The descriptive statistics of daily log returns for the $39$ companies of our sample is presented in Table \ref{tab:stats}, Appendix \ref{app:daily returns}.

\begin{figure}[h]
    \centering

    \begin{subfigure}[b]{0.49\textwidth}
        \centering
        \includegraphics[width=\textwidth]{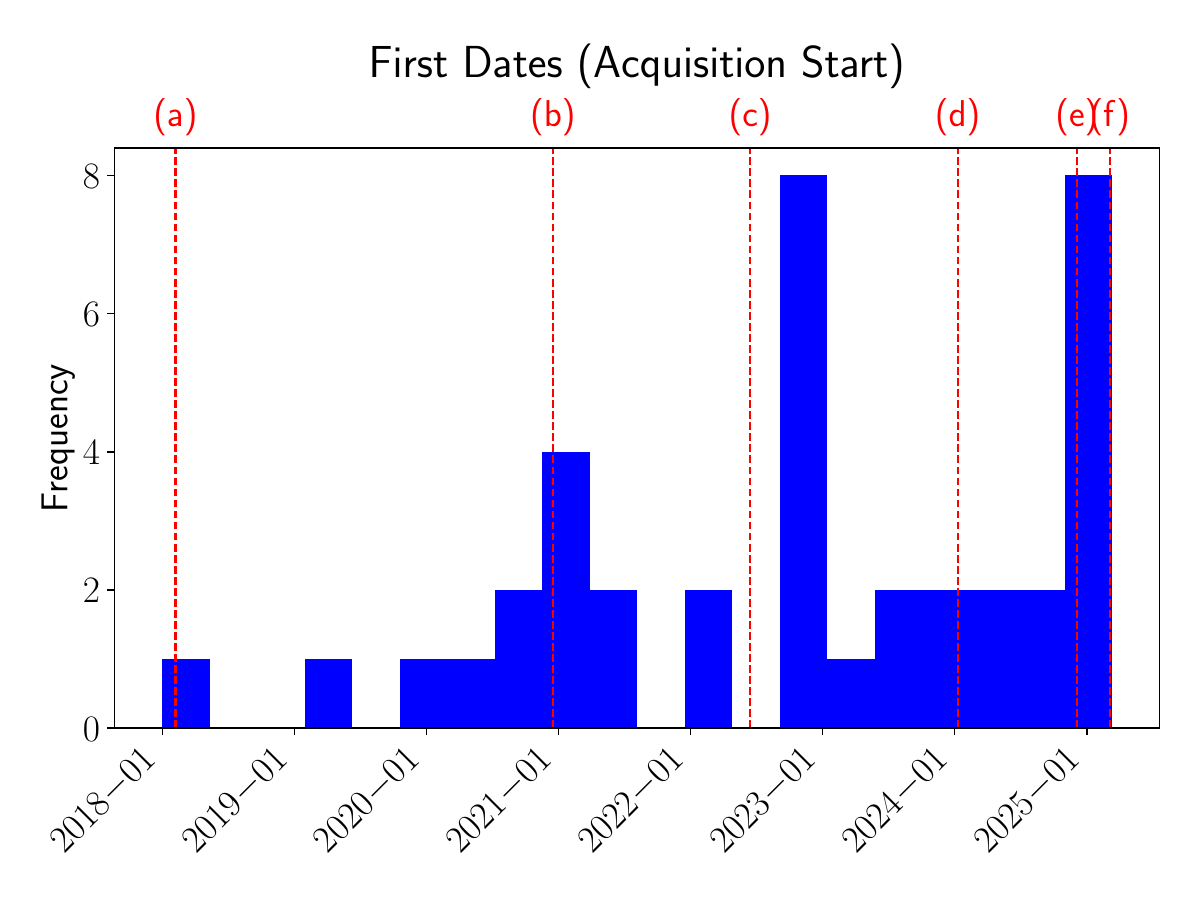}
        \caption{}
        \label{fig:first_dates}
    \end{subfigure}
    %\hfill
    \begin{subfigure}[b]{0.49\textwidth}
        \centering
        \includegraphics[width=\textwidth]{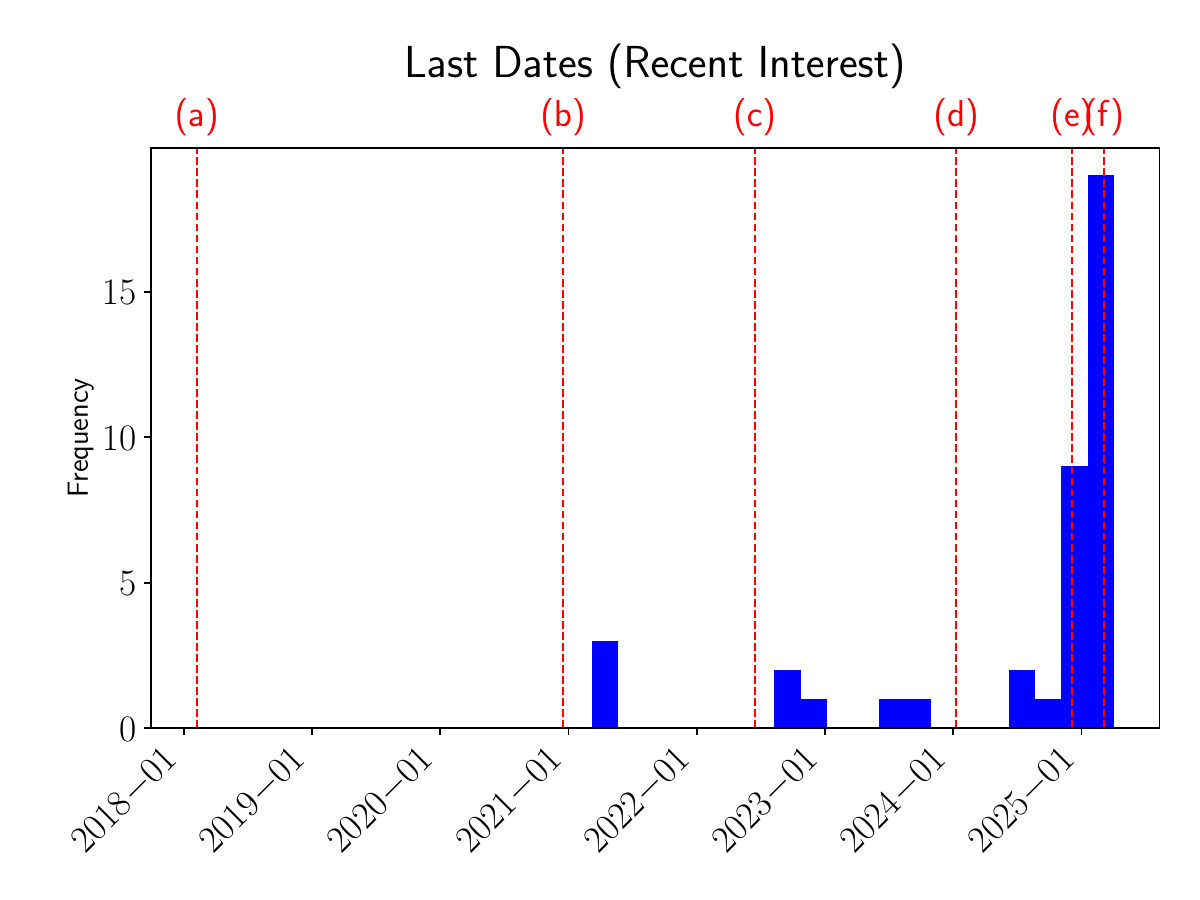}
        \caption{}
        \label{fig:last_dates}
    \end{subfigure}

    %\vspace{1em} % vertical spacing between rows

    % Second row: centered single subfigure
    \begin{subfigure}[b]{0.49\textwidth}
        \centering
        \includegraphics[width=\textwidth]{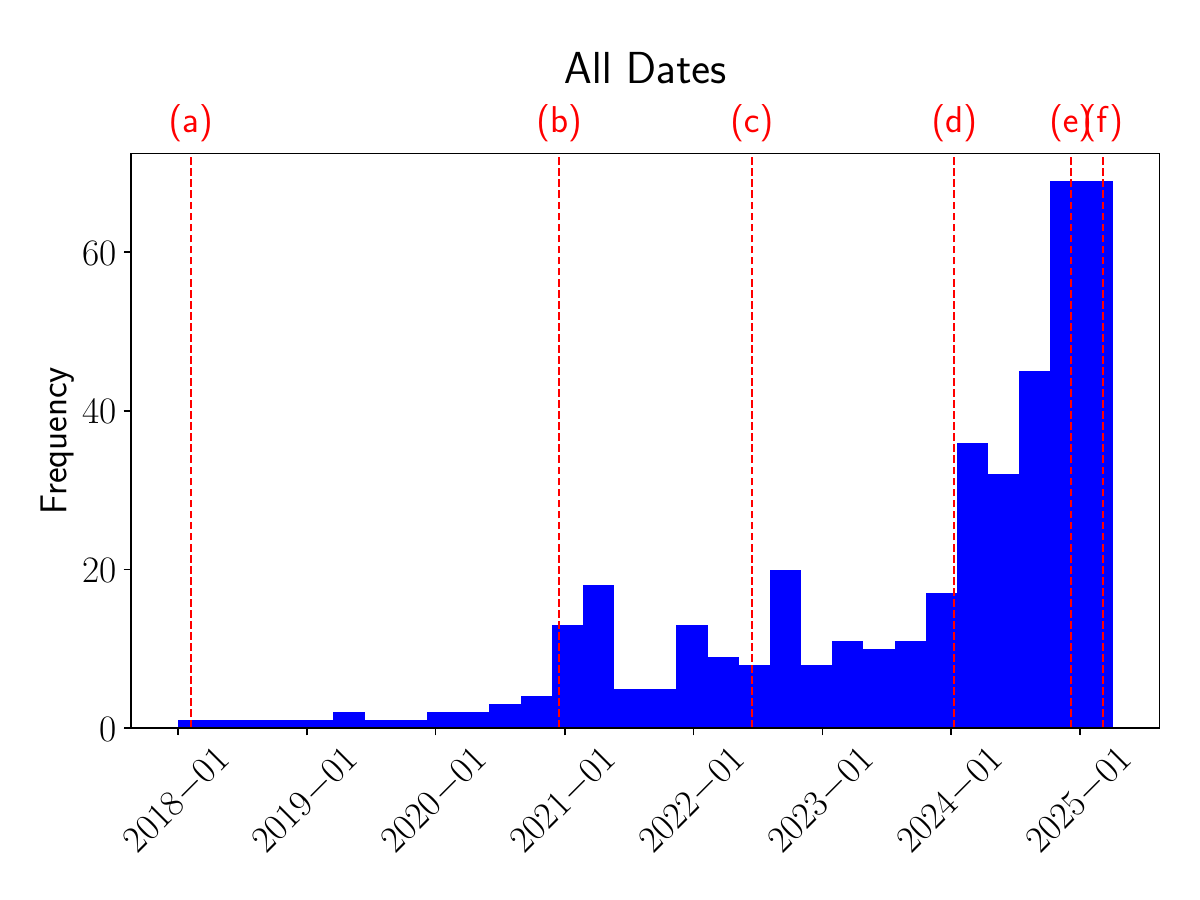}
        \caption{}
        \label{fig:all_dates}
    \end{subfigure}

    \caption{Red vertical lines mark key events in the BTC market timeline. Figure \ref{fig:first_dates}: distribution of first acquisition dates across companies in the dataset, highlighting the initial waves of institutional adoption. Figure \ref{fig:last_dates}: distribution of last acquisition dates, capturing more recent activity and assessing whether interest is currently accelerating. Figure \ref{fig:all_dates}: overall distribution of BTC acquisition dates for all recorded transactions in the dataset.
     There is a clear acceleration in acquisition activity starting in early $2024$, peaking in $2025$. Earliest first date: Dec $31$, $2017$ (\textit{Hut 8 Mining Corp}); latest last date: April $3$, $2025$ (\textit{MARA Holdings, Inc.} and \textit{Riot Platforms, Inc.}). Data are updated as of April $11$, $2025$.}
    \label{fig:dates_acquisition}
\end{figure}

Figure \ref{fig:dates_acquisition} offers a comprehensive overview of the temporal distribution of institutional BTC acquisitions. Panel \ref{fig:first_dates} details the timing of initial acquisitions, revealing that early institutional involvement between $2018$ and $2022$ was both limited and sporadic, likely reflecting cautious exploratory behavior in a context of regulatory uncertainty and market volatility. A discernible increase in first time acquisitions begins in $2021$, coinciding with the post-pandemic bull market, heightened media attention, and a growing recognition of BTC as a potential hedge against inflation and currency debasement. The most pronounced surge, however, emerges in $2023$, followed by a secondary peak in early $2025$. This acceleration suggests successive waves of institutional entry, potentially catalyzed by the approval of spot BTC exchange-traded funds, the stabilization of global monetary policy, and a post U.S. election market rally in late $2024$.
Panel \ref{fig:last_dates}, which depicts the distribution of final acquisition dates, corroborates this trajectory. While earlier years exhibit fragmented activity, the data show a clear concentration of last acquisitions in $2024$ and early $2025$. This pattern indicates sustained institutional accumulation, reflecting strategic long term positioning and growing confidence in BTC’s market legitimacy. Contributing factors likely include improved regulatory clarity, custodial and trading infrastructure maturation, and evolving macroeconomic conditions favoring alternative assets.
Panel \ref{fig:all_dates} aggregates all acquisition dates into a unified histogram, providing a view of the temporal dynamics of BTC purchases. The frequency of acquisitions increased markedly beginning in early $2024$, peaking in $2025$. Whereas earlier institutional activity was sparse and opportunistic, the recent surge emphasise a broad-based shift toward BTC adoption. This shift is driven not only by favorable regulatory developments, such as ETF approvals, but also by institutional portfolio diversification strategies, increased liquidity, and the perceived resilience of BTC among global financial uncertainties.

The red vertical lines in Figure \ref{fig:dates_acquisition} represent significant events in the history of BTC markets, happened in the period spanning from December $2018$ to April $2025$:

\begin{enumerate}[label=\textbf{(\alph*)}]
    \item \textbf{February $6$, $2018$ – BTC Price Correction}: After reaching a peak of approximately $\$19,000$ in November $2017$, BTC's price experienced a sharp decline, falling below $\$6,000$ by early February $2018$. From January $6$, $2017$, to February $6$, $2017$, the cryptocurrency lost roughly $65\%$ of its value, marking one of the most severe drawdowns in its history following the unprecedented 2017 bull market \cite{bbc2018bitcoin}.
    \item \textbf{December $16$, $2020$ – Post COVID Rally}: Following a low of around $\$3,600$ in March $2020$, triggered by the global market sell-off coinciding with the COVID-19 pandemic, BTC's price surged by over $400\%$ throughout the year. On December $16$, $2020$, it surpassed $\$20,000$ for the first time, reaching a new all-time high and marking the beginning of another major bull cycle \cite{partington2020bitcoin}.
    \item \textbf{June $15$, $2022$ – Historic Monthly Loss}: BTC experienced its most severe monthly decline in over a decade, falling by $37.3\%$ in June $2022$. This drop represented the largest monthly loss since $2011$ and occurred in correspondence of a broader market downturn \cite{he2022brutal}.
    \item \textbf{January $10$, $2024$ – SEC Approves BTC ETFs}: The U.S. SEC approved $11$ BTC ETF applications from major financial institutions, including BlackRock, Fidelity, and VanEck. This regulatory breakthrough expanded institutional access to BTC via traditional investment channels \cite{yerushalmy2024bitcoinetf}.
    \item \textbf{December $5$, $2024$ – Post U.S. Election Rally}: Following the outcome of the U.S. presidential election, BTC's price surged past $\$100,000$ for the first time. The rally was driven by investor expectations of more favorable regulatory conditions under the new administration \cite{ray2024bitcoin}.
    \item \textbf{March $6$, $2025$ - The White House commitment}: U.S. President Donald Trump signed an executive order establishing a Strategic Bitcoin Reserve and a U.S. Digital Asset Stockpile. The move framed BTC as a sovereign reserve asset \cite{whitehouse2025bitcoinreserve}.
\end{enumerate}

\begin{figure}[h]
    \centering
    \includegraphics[width=\linewidth]{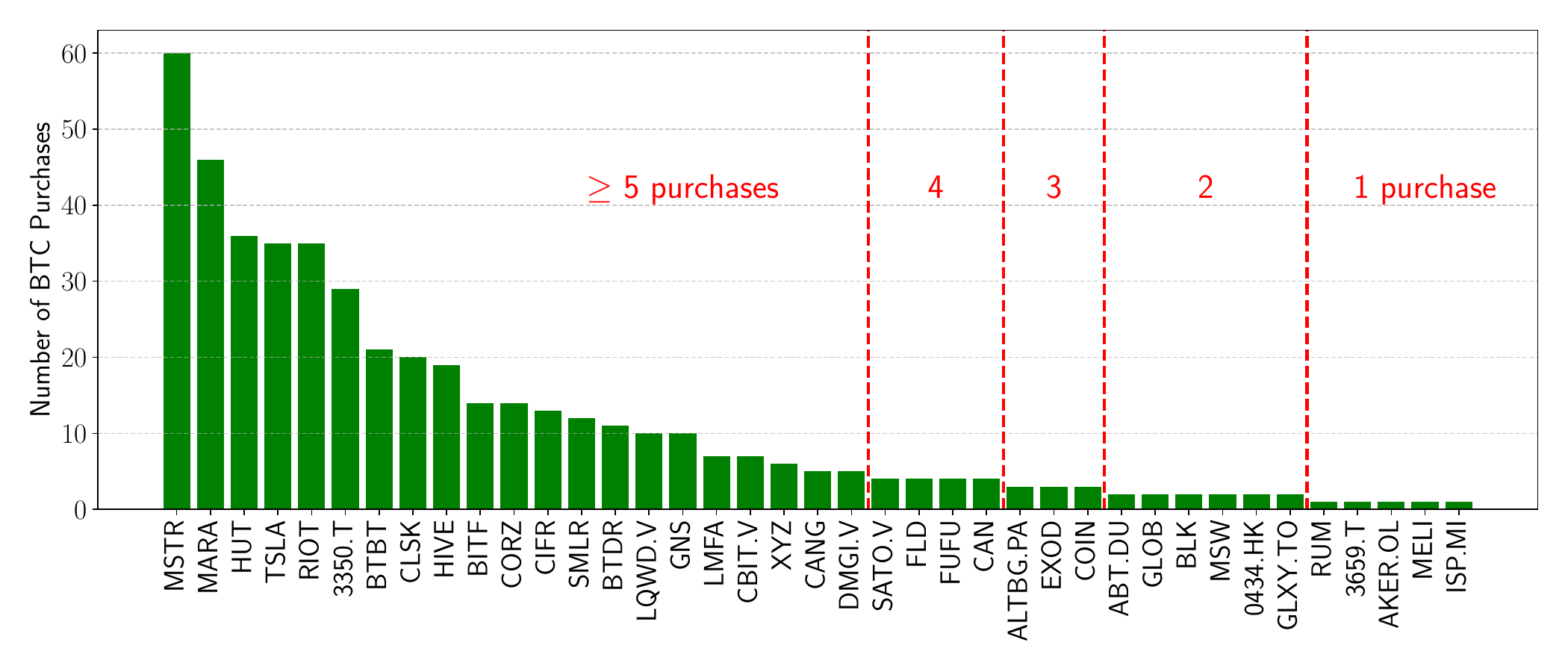}
    \caption{Frequency of BTC purchased by firm.}
    \label{fig:number purchases}
\end{figure}

\begin{figure}
    \centering
    \includegraphics[width=0.5\linewidth]{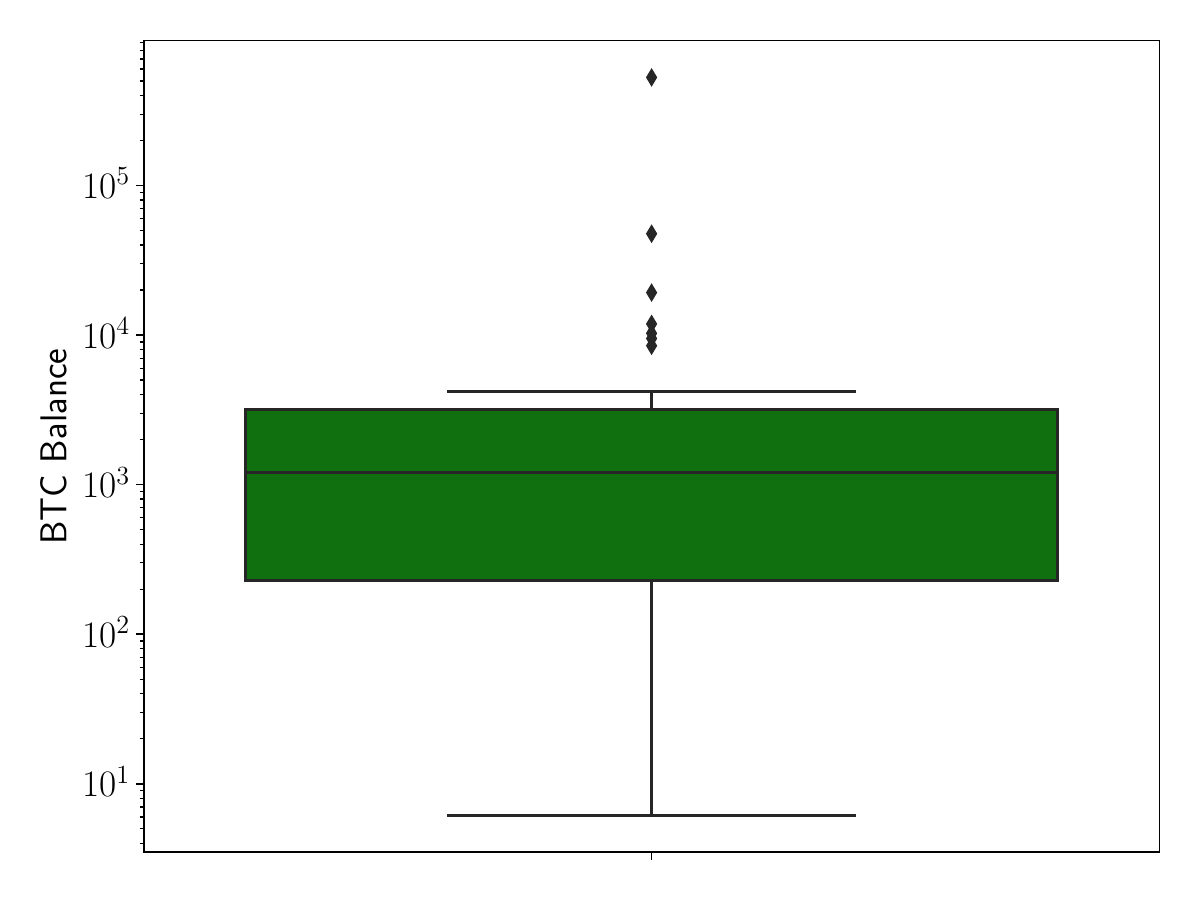}
    \caption{The box plot displays the distribution of BTC holdings among the $39$ companies in our dataset. The lower and upper edges of the box correspond to the first ($25\%$) and third ($75\%$) quartiles, respectively; the line within the box marks the median ($50\%$).  Whiskers extend to the minimum and maximum values within the distribution, and small diamond markers beyond the whisker indicate outlier firms with exceptionally large BTC positions.}
    \label{fig:btc balance}
\end{figure}

Figure \ref{fig:number purchases} summarises the intensity of BTC acquisition activity across the $39$ companies in our sample: $5$ of them executed only a single purchase, and $13$ conducted two to four discrete transactions. The majority ($21$ of them) -- including several crypto-native miners -- undertook five or more separate purchases, signalling an active approach to balance sheet accumulation. Figure \ref{fig:btc balance} complements this by illustrating the distribution of BTC holdings via a box plot. The lower and upper edges of the box represent the $25$th ($228$ BTC) and $75$th ($3185$ BTC) percentiles, respectively, with the central line marking the median holding ($1200$ BTC). Whiskers extend to the minimum ($6.15$ BTC held by \textit{BlackRock, Inc.}) and maximum ($528,185$ BTC held by \textit{Strategy}) values, and diamonds denote outliers beyond these bounds. The pronounced skew of the distribution highlights that a small number of firms -- most notably \textit{Strategy} -- hold large BTC positions compared to the rest of the sample.

\subsection{\textit{Strategy} (MSTR)}

Since this paper centers on \textit{Strategy}, the largest corporate holder of BTC, we confine our detailed analysis exclusively to this company.

\begin{figure}[h]
    \centering
    \includegraphics[width=\linewidth]{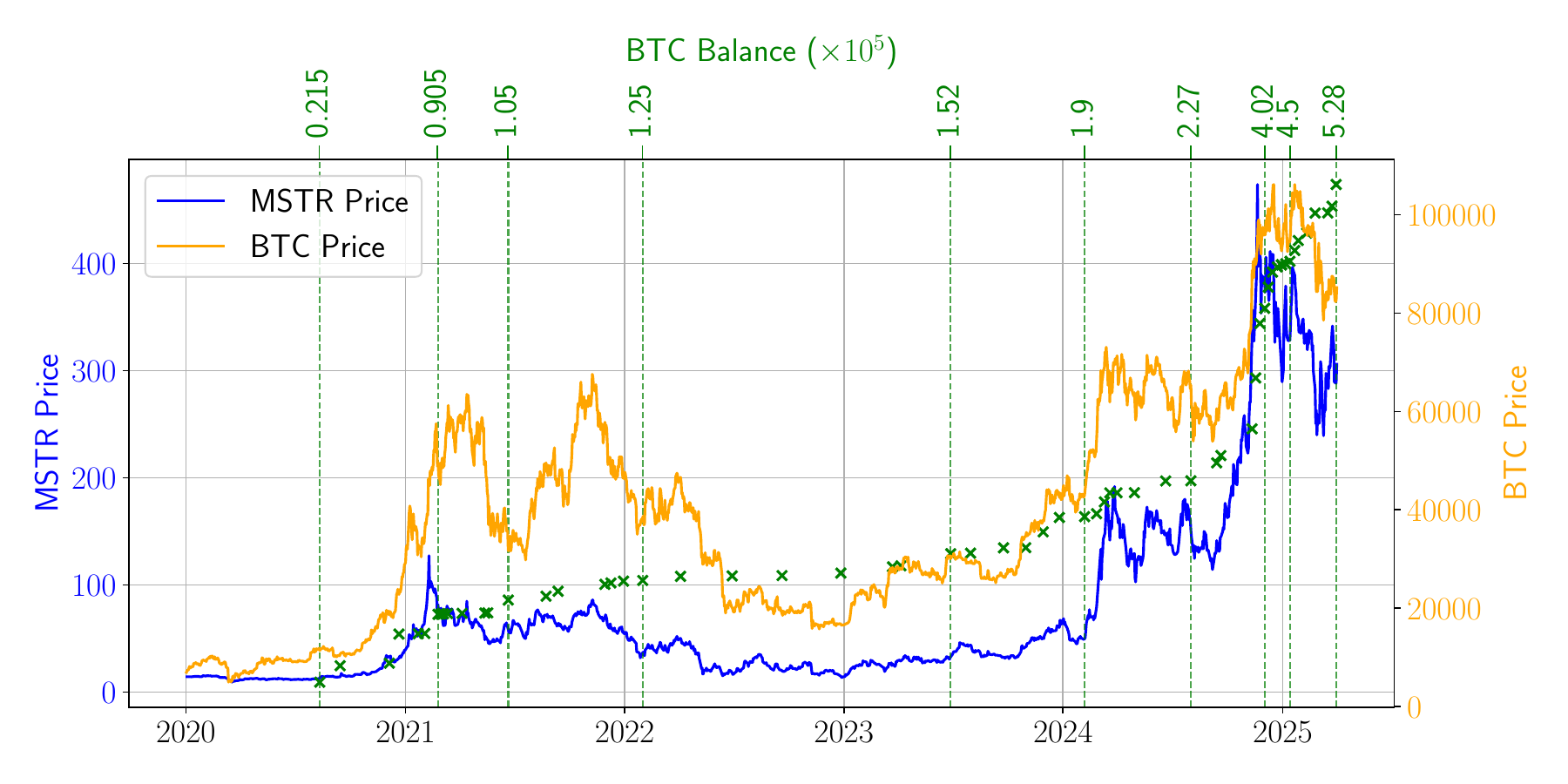}
    \caption{BTC price (in orange), MSTR price (in blue), and cumulative BTC holdings (in green).}
    \label{fig:MSTR_descriptive}
\end{figure}

Following its most recent purchase on March $31$, $2025$, the company’s cumulative BTC holdings reached $528,185$ BTC. Figure \ref{fig:MSTR_descriptive} illustrates the relationship between BTC acquisitions and daily market dynamics.  As of April $2025$, \textit{Strategy}’s equity has delivered a $21\%$ total return over the previous three months and a $229\%$ return over the past year.  Its implied volatility is $72\%$, indicating that market participants expect future price fluctuations of this magnitude throughout the life of the outstanding MSTR options; indeed, the $30$ day historical volatility, calculated as the standard deviation of daily log returns over the last month, measures $59\%$. Table \ref{tab:stats_MSTR_BTC} reproduces only the rows from Table \ref{tab:stats} (in Appendix \ref{app:daily returns}) that pertain to MSTR and BTC, presenting a descriptive statistics of daily log returns from April $1$, $2023$ to April $1$, $2025$.

\begin{table}[h]
\centering
\begin{tabular}{lccccccc}
\toprule
\textbf{Ticker} & \textbf{Mean} & \textbf{Median} & \textbf{Std} & \textbf{Skewness} & \textbf{Kurtosis} & \textbf{Min.} & \textbf{Max.} \\
\midrule
BTC-USD & 0.0015 & 0.0002 & 0.0252 & 0.3056 & 2.3004 & -0.0908 & 0.1146 \\
MSTR    & 0.0047 & 0.0007 & 0.0582 & 0.1576 & 1.5407 & -0.2384 & 0.2290 \\
\bottomrule
\end{tabular}
\caption{Descriptive statistics of daily log returns for BTC and MSTR, from April 1, 2023 to April 1, 2025.}
\label{tab:stats_MSTR_BTC}
\end{table}

From $2021$ through $2024$, MSTR stock price significantly outperformed BTC, growing $7.5\times$ versus $3.2\times$.
Between late August and late October 2024, MSTR’s share price nearly doubled, whereas BTC rose by only about $24\%$. 
The divergence is especially pronounced over 5 to 7 October 2024, when MSTR gained $18\%$ while BTC advanced by just $1\%$.
According to Phillips \& Pohl (2024) \cite{phillips2024microstrategy}, this decoupling stems from MSTR evolving strategy, notably its rebranding as a ``Bitcoin Treasury Company''. This shift involved aggressive capital raising -- issuing equity and debt -- to fund substantial BTC acquisitions, thereby enhancing its leverage and altering its capital structure. 
%The company introduced the ``Bitcoin Yield'' metric to quantify the accretive impact of these strategies on shareholder value. This approach attracted attention from both practitioners and academics interested in complete markets and corporate finance dynamics, as some investors are prevented from buying Bitcoin directly. Therefore, MSTR works as a synthetic instrument for investors interested in Bitcoin exposure. This generates an increase in demand for MSTR's shares above its fundamental value, since it grants access to an otherwise inaccessible asset. 
Choueifaty et al. (2025) \cite{choueifaty2025accounting} provide a decomposition of the MSTR performance into three sources: (i) monetization of a significant portion of the pre-existing premium into book value, (ii) leveraging BTC's performance through increased exposure, and (iii) fluctuations in the BTC price premium.
Notice that the public disclosure of \textit{Strategy}’s BTC acquisitions is almost immediate, as the firm typically files a Form 8-K and issues a press release within hours, usually by the next trading session, of completing each purchase \cite{microstrategy2025form8k}.

\section{Methodology}\label{sec:Methodology}

To probe the directional interdependencies between daily BTC returns and the equity returns of the firms in our sample, we estimate the Transfer Entropy (TE), an information measure that captures lagged, conditional dependence between time series. Let  
\begin{equation}
\mathbf{X}_{t,\boldsymbol{\tau}}^{-} = \bigl(X_{t-\tau_{1}},\dots,X_{t-\tau_{j}}\bigr), \qquad   
\mathbf{Y}_{t,\boldsymbol{\lambda}}^{-} = \bigl(Y_{t-\lambda_{1}},\dots,Y_{t-\lambda_{k}}\bigr)
\end{equation}
denote, respectively, the vectors containing the past $j$ lags of the BTC return series $X_t$ and the past $k$ lags of a given firm’s return series $Y_t$. Under the Wiener-Granger causality framework \cite{granger1969investigating}, $X_t$ is said to \emph{cause} $Y_t$ at prediction horizon $h$ if the history of $X_t$ improves forecasts of $Y_{t+h}$ beyond what is achievable using only the past of $Y_t$. As stated in Equation \ref{eq:transfer_entropy_difference}, the TE from $X$ to $Y$ at horizon $h$ is defined as the reduction in uncertainty, quantified by Shannon entropy $H(\cdot)$, about $Y_{t+h}$ when including the history of $X_t$ in the conditioning set.

\begin{equation} \label{eq:transfer_entropy_difference}
    T_{\mathbf{X} \rightarrow \mathbf{Y}} = H\left( Y_{t+h} \bigm| \mathbf{Y}_{t, \boldsymbol{\lambda}}^{-} \right) - H \left( Y_{t+h} \bigm| \mathbf{X}_{t, \boldsymbol{\tau}}^{-}, \mathbf{Y}_{t, \boldsymbol{\lambda}}^{-} \right)
\end{equation}
\vspace{0.05cm}

By applying the definition for the conditional mutual information measure (i.e., $I(A;B|C) = H(A|C) - H(A|B,C)$), Equation \ref{eq:transfer_entropy_difference} can be written as per in Equation \ref{eq:transfer_entropy_mi}.

\begin{equation}\label{eq:transfer_entropy_mi}
    T_{\mathbf{X} \rightarrow \mathbf{Y}} = I\bigl( Y_{t+h} ; \mathbf{X}_{t, \boldsymbol{\tau}}^{-} \bigm| \mathbf{Y}_{t, \boldsymbol{\lambda}}^{-} \bigr)
\end{equation}

Formally, the conditional mutual information can be expressed as an expectation over the joint distribution of the relevant variables:
\begin{equation}
T_{\mathbf{X} \rightarrow \mathbf{Y}} = \mathbb{E} \left[ \log \frac{ P(Y_{t+h} \mid \mathbf{X}_{t, \boldsymbol{\tau}}^{-}, \mathbf{Y}_{t, \boldsymbol{\lambda}}^{-}) }{ P(Y_{t+h} \mid \mathbf{Y}_{t, \boldsymbol{\lambda}}^{-}) } \right],
\end{equation}
where $P(\cdot)$ denotes the appropriate conditional probability distributions. This highlights that transfer entropy measures the expected information gain, in bits, from incorporating the history of $X_t$ into predictions of $Y_{t+h}$.

By construction, $T_{\mathbf{X} \rightarrow \mathbf{Y}} \geq 0$, with $T_{\mathbf{X} \rightarrow \mathbf{Y}} = 0$ indicating no causal influence from $X$ to $Y$ at the specified lags. To assess whether an observed value of transfer entropy reflects a genuine directional dependency rather than random fluctuations, a non-parametric hypothesis testing procedure is employed. The null hypothesis $H_0$ assumes no directional influence from $X$ to $Y$, meaning any observed transfer entropy arises purely by chance. To approximate the distribution of $T_{\mathbf{X} \rightarrow \mathbf{Y}}$ under $H_0$, surrogate datasets are generated by independently reshuffling the time indices of the driver series $X_t$. This procedure preserves the marginal distribution of $X_t$ but destroys its temporal and structural dependencies with $Y_t$. For each shuffled sample $i = 1, \dots, N^{\text{shuffle}}$, the corresponding transfer entropy value $T^{\text{shuffle}}_i$ is computed, yielding an empirical null distribution. The significance of the observed transfer entropy $T_{\mathbf{X} \rightarrow \mathbf{Y}}$ is evaluated by calculating the proportion of shuffled values that exceed or equal it:
\begin{equation}
p = \frac{1}{N^{\text{shuffle}}} \sum_{i=1}^{N^{\text{shuffle}}} \mathbb{I}\left( T^{\text{shuffle}}_i \geq T_{\mathbf{X} \rightarrow \mathbf{Y}} \right),
\end{equation}
where $\mathbb{I}(\cdot)$ is the characteristic function. This empirical $p$-value estimates the probability of observing a transfer entropy as large as or larger than the measured value under the null hypothesis. A small $p$-value indicates that the observed directional dependency is unlikely to have occurred by chance, providing evidence against $H_0$ and supporting the presence of a meaningful causal relationship.

\section{Results}\label{sec:Results}

Guided by the temporal trend shown in Fig. \ref{fig:all_dates}, and supported by the empirical analyses presented in Sections \ref{sec:global measures} and \ref{sec:sfm}, we focus on the most recent two-year period, from April $1$, $2023$ to April $1$, $2025$ ($\sim500$ trading days). The code for reproducing all experiments is available at \url{https://github.com/FinancialComputingUCL/Crypto_Balance_Sheets}.

\subsection{Global Measures of Information}\label{sec:global measures}

In our preliminary analysis, we compute three Pearson correlation (PC) coefficients to examine both same-day and lagged one-day return dependencies between BTC and every stock \( X \) in the sample:

\begin{enumerate}
    \item \(\rho(BTC_t, X_t)\): the same-day correlation between BTC and equity returns;
    \item \(\rho(BTC_t, X_{t-1})\): the correlation capturing the extent to which BTC returns anticipate equity returns by one day;
    \item \(\rho(BTC_{t-1}, X_t)\): the correlation assessing whether equity returns precede BTC returns by one day.
\end{enumerate}

Analyzing lagged correlations helps identify potential lead-lag dynamics that may arise from information transmission delays, market microstructure effects, or behavioral asymmetries. Such dynamics are closely related to the concept of directional dependence, which we further explore using TE\footnote{TE estimates are obtained with the R package `RTransferEntropy' \cite{behrendt2019rtransferentropy}.}. Indeed, to quantify the direction and magnitude of information flow, we integrate these correlation measures with two TE estimates:

\begin{enumerate}[start=4]
    \item \( TE_{BTC(t-1) \rightarrow X(t)} \): the information transfer from BTC returns at lag $1$ to current stock $X$ returns;
    \item \( TE_{X(t-1) \rightarrow BTC(t)} \): the information transfer from stock $X$ returns at lag $1$ to current BTC returns.
\end{enumerate}

{\tiny
\begin{longtable}[c]{@{}lccccc@{}}
\toprule
\textbf{Statistic} & 
\textbf{\begin{tabular}[c]{@{}c@{}}PC: \\ BTC (t) $\rightleftarrows$ X(t)\end{tabular}} &
  \textbf{\begin{tabular}[c]{@{}c@{}}PC: \\ BTC (t) $\rightleftarrows$ X(t-1)\end{tabular}} &
  \textbf{\begin{tabular}[c]{@{}c@{}}PC: \\ BTC (t-1) $\rightleftarrows$ X(t)\end{tabular}} &
  \textbf{\begin{tabular}[c]{@{}c@{}}TE: \\ BTC(t-1) $\rightarrow$ X(t)\end{tabular}} &
  \textbf{\begin{tabular}[c]{@{}c@{}}TE: \\ X(t-1) $\rightarrow$ BTC(t)\end{tabular}} \\*
\midrule
\endfirsthead
\multicolumn{6}{c}{{\bfseries Table \thetable\ continued from previous page}} \\
\toprule
\textbf{Statistic} & 
\textbf{PC: BTC(t) $\rightleftarrows$ X(t)} & 
\textbf{PC: BTC(t) $\rightleftarrows$ X(t-1)} & 
\textbf{PC: BTC(t-1) $\rightleftarrows$ X(t)} & 
\textbf{TE: BTC(t-1) $\rightarrow$ X(t)} & 
\textbf{TE: X(t-1) $\rightarrow$ BTC(t)} \\
\midrule
\endhead
Mean     & 0.292 & -0.018 & 0.018  & 0.0151 & 0.0144 \\
Median   & 0.254 & -0.014 & 0.004  & 0.0140 & 0.0140 \\
Std      & 0.209 & 0.040  & 0.0782 & 0.0066 & 0.0060 \\
Skewness & 0.164 & -1.105 & 1.780  & 0.176  & 0.787  \\
Kurtosis & -1.363 & 0.893 & 4.906  & 0.170  & 1.186  \\
\bottomrule
\caption{Summary statistics of the dependence measures between BTC and the 39 stocks. The table reports the mean, median, standard deviation, skewness, and kurtosis for each of the five global information metrics discussed earlier in the section.}
\label{tab:compact_results}
\end{longtable}}

Extended results are reported in Table~\ref{tab:table_correlations} in Appendix~\ref{app:correlation}, while a concise yet informative summary is provided in Table \ref{tab:compact_results}. The mean $\rho(BTC_t, X_t)$ is moderately positive (i.e., $0.29$), indicating that, on average, stocks co-move with BTC, likely reflecting shared exposure to the digital asset ecosystem. The relatively low standard deviation (i.e., $0.21$) and mild skewness (i.e., $0.16$) suggest a fairly symmetric distribution centered around positive values. The negative kurtosis reflects a light tailed distribution, implying few extreme correlation values. In contrast, $\rho(BTC_t, X_{t-1})$ and $\rho(BTC_{t-1}, X_t)$ are centered near $0$, with means of $-0.018$ and $0.018$, respectively. This indicates that no consistent linear lead-lag pattern emerges across the sample. However, their skewness values diverge: BTC-leading correlations are negatively skewed (i.e., $-1.11$), suggesting that a subset of stocks exhibit pronounced negative lagged correlation with future BTC returns; while equity-leading correlations are positively skewed (i.e., $1.78$), with significant excess kurtosis (i.e., $4.91$), pointing to heavy tails. This hints at asymmetric or nonlinear relationships, where a few stocks might carry signal that anticipate BTC, even if the overall mean effect is muted. These patterns motivate the use of transfer entropy (TE), which can capture nonlinear and directional information flows. The TE results show low mean values ($0.0151$ from BTC to stocks, $0.0144$ in the reverse direction), but with positive skewness and mild kurtosis, especially for $TE_{X(t-1) \rightarrow BTC(t)}$.  Consistently with isolated lead-lag relationships observed in the correlation tails, this suggests that while the average information transfer is limited, a subset of equities might sporadically transmit information to BTC.

While these global measures offer valuable insights, they also smooth over temporal variation that may be crucial for understanding dynamic relationships among financial assets. The asymmetric distribution of lagged correlations and the episodic nature of significant transfer entropy values suggest that informational dependencies between BTC and equities are highly unstable over time. As such, relying on full-sample estimates may obscure localized periods of directional influence or structural shifts in market behavior. To uncover these dynamics, in Section \ref{sec:rolling_transfer_entropy}, we proceed with  a rolling estimation of the TE, which enables us to capture time-varying information flow and identify windows where BTC or equity returns exert a statistically significant influence on one another. This local analysis is especially pertinent for real-time forecasting, trading strategies, and market surveillance, where detecting regime shifts and lead-lag transitions is more informative than long-run averages.

\subsection{Single Factor Model: Exposure Driven BTC $\beta$}\label{sec:sfm}

To assess the sensitivity of equities to BTC price fluctuations, we estimate a single factor model \cite{fama1993common} for each firm $i$ in the sample:

\begin{equation}\label{eq:regression}
r_{i, t} = \alpha_i + \beta_i r_t^{\mathrm{BTC}} + \varepsilon_{i, t}, \quad t = 1, \ldots, T
\end{equation}

where $r_{i, t}$ and $r_t^{\mathrm{BTC}}$ denote the daily log returns of stock $i$ and BTC, respectively; $\alpha_i$ is the firm-specific intercept; and $\beta_i$ represents the stock’s sensitivity to BTC returns, that is, the percentage change in the stock's return associated with a 1\% change in BTC on the same trading day.
Companies with \(\beta>1\) are characterized by leveraged exposure to BTC price changes, while those with \(\beta\le 1\) exhibit proportional or lower sensitivity. We report the regression coefficients for the whole dataset in Table \ref{tab:tab betas} in Appendix \ref{app:sfm}, highlighting that $12$ firms exhibit a $\beta > 1$, while the average beta is $\hat\beta = 0.62$. 

\begin{figure}[h]
    \centering
    \includegraphics[width=0.75\linewidth]{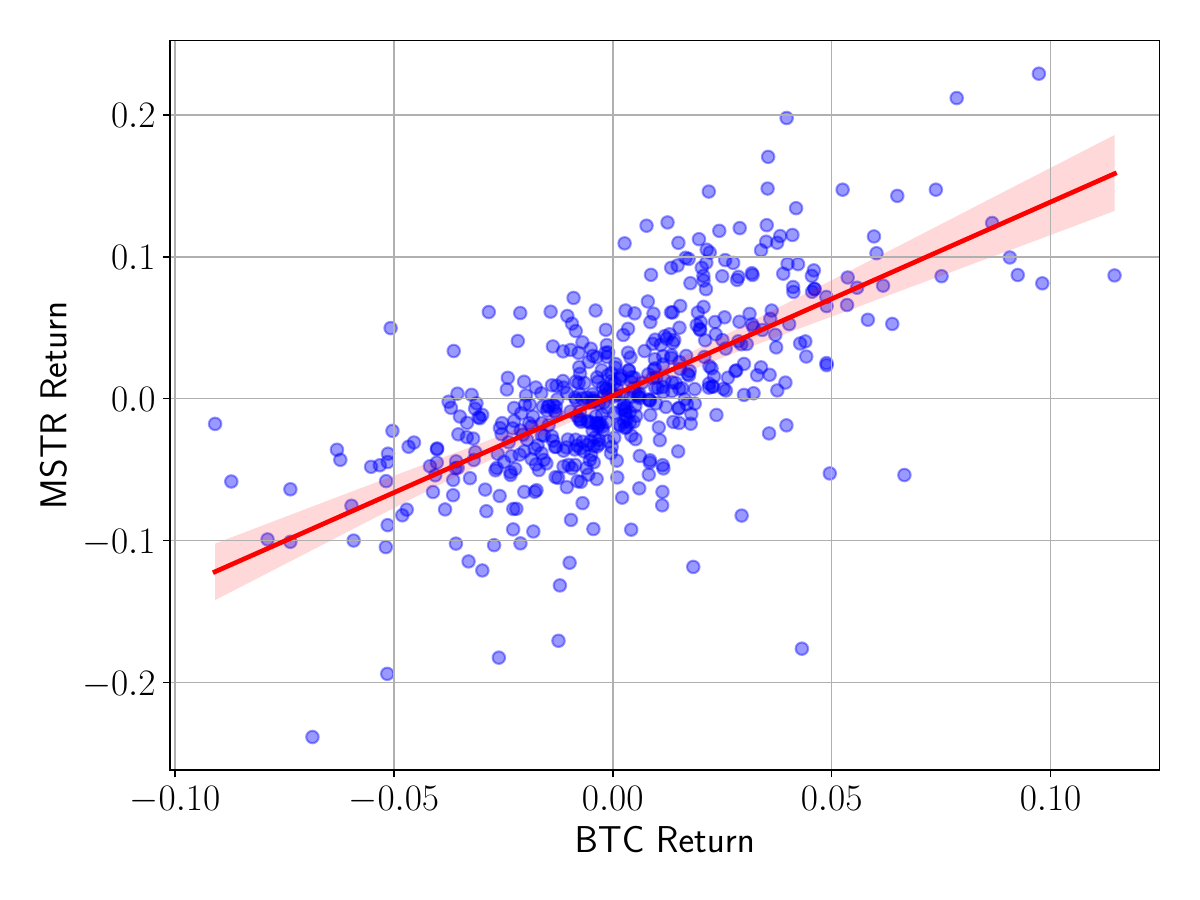}
    \caption{Single factor model of MSTR daily returns on BTC (Apr $1$, $2023$ to Apr $1$, $2025$). The estimated BTC $\beta = 1.37$ indicates that a $1\%$ move in BTC is associated with an average $1.37\%$ move in MSTR; $\alpha = 0.0020$; $R^{2}=0.44$. 
    The red band depicts the $99\%$ confidence interval around the fitted line.}
    \label{fig:single fm mstr}
\end{figure}

Figure \ref{fig:single fm mstr} reports the results from a single factor model regression for MSTR daily log returns on BTC ones over the period spanning from April $1$, $2023$, to April $1$, $2025$. The estimated factor loading ($\beta = 1.37$) implies that a $1$\% increase in BTC returns is associated with an average $1.37$\% increase in MSTR returns, consistent with a high degree of systematic exposure to BTC price risk. The model explains a substantial proportion of return variation, with an $R^2$ of $0.44$, indicating that BTC returns account for $44$\% of the variance in MSTR’s daily returns.

\begin{figure}[h]
    \centering
    \includegraphics[width=0.9\linewidth]{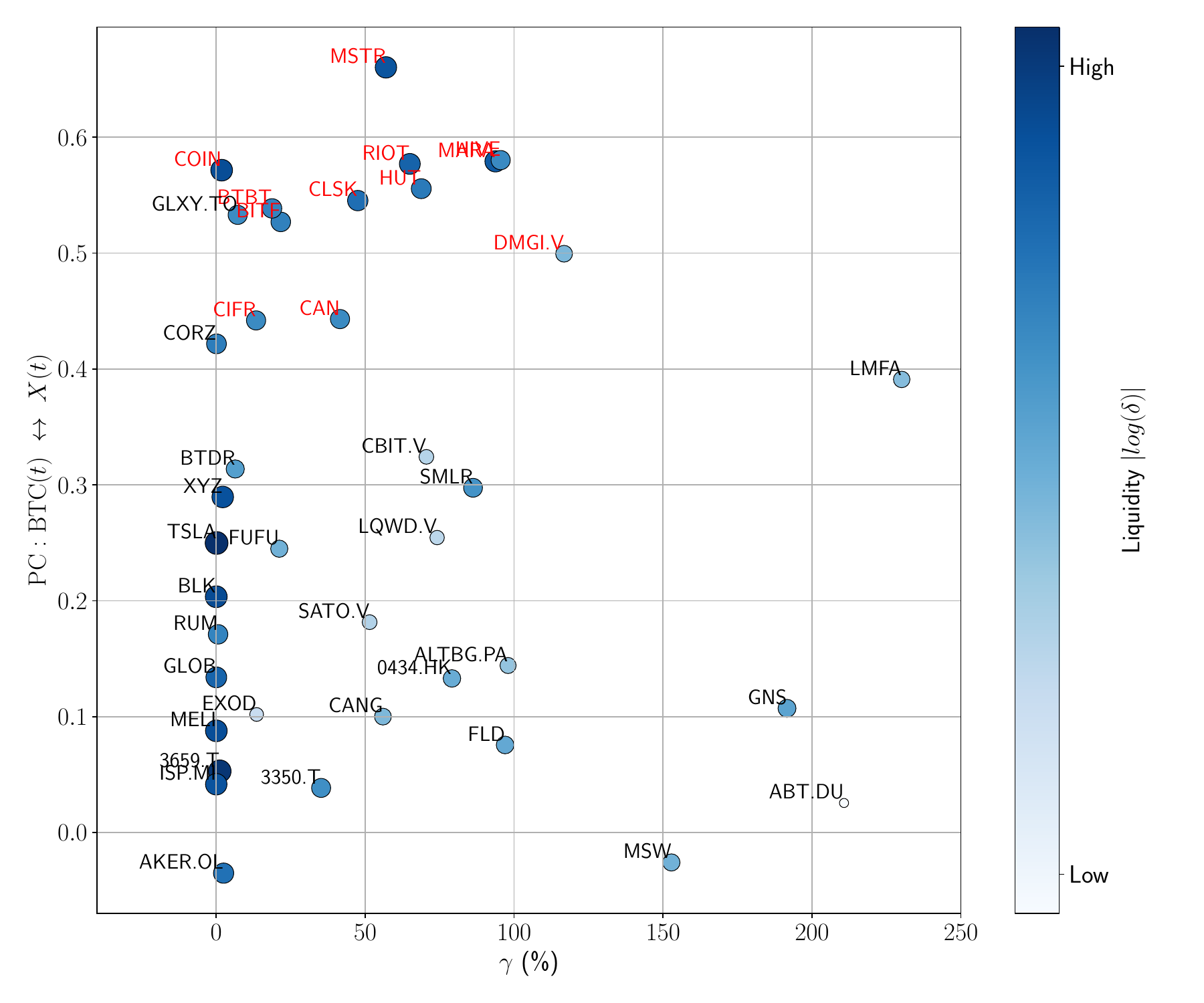}
    \caption{BTC and equity return correlations with exposure and liquidity indicators for $39$ firms. $x‑$axis: $\gamma$, defined as BTC holdings as a percentage of market capitalisation (at the latest acquisition date). $y‑$axis: Pearson correlation between same‑day stock and BTC returns (Apr $1$, $2023$ - Apr $1$, $2025$).  Marker size and colour are proportional to $|\log(\delta)|$: darker, larger markers correspond to more liquid stocks. Red tickers denote firms with $\beta>1$; black tickers denote $\beta\le 1$.}
    \label{fig:btc-marketcap}
\end{figure}

Figure \ref{fig:btc-marketcap} combines two global measures: the Pearson correlation (discussed in Section \ref{sec:global measures}) and the estimated betas. The $x$-axis reports each firm’s BTC intensity $\gamma$, measured as BTC holdings expressed as a percentage of market capitalization, while the $y$-axis shows the Pearson correlation between the firm’s daily log returns and BTC returns. Firms with $\beta > 1$ are labeled in red, indicating high sensitivity to BTC price movements, whereas those with $\beta \leq 1$ are labeled in black. Marker size and color gradient are proportional to the Amihud illiquidity ratio $\delta$ \cite{amihud2002illiquidity} in Equation \ref{eq:amihud}, defined as the daily ratio of a stock’s absolute return to its dollar trading volume. This ratio reflects the daily price response to one dollar of trading volume and serves as an approximate measure of price impact.

\begin{equation}\label{eq:amihud}
    \delta =\frac{1}{N}\sum_{t=1}^{N}\frac{|\,r_{t}\,|}{\text{DollarVolume}_{t}},
\end{equation}

The original Amihud ratio $\delta$ (where $0<\delta<1$) measures price impact: a high $\delta$ implies that even small trade volumes cause large price moves (low liquidity), whereas a low $\delta$ indicates that larger volumes are required to move the price (high liquidity). To invert this scale and align larger values with greater liquidity, we use $|\log(\delta)|$. Under this transformation, high values of $|\log(\delta)|$ correspond to original $\delta \sim 0$, that is more liquid stocks; while low $|\log(\delta)|$ values coincide with illiquid stocks.  Consequently, in our plot, darker and larger markers (high $|\log(\delta)|$) denote highly liquid equities, whereas lighter and smaller markers indicate lower liquidity.
It is worth emphasizing that the Amihud illiquidity ratio is not a conventional liquidity measure, such as the bid–ask spread or market depth. Instead, it captures price impact, measuring the average return required to absorb one dollar of trading volume. A lower ratio implies that sizable trades can be executed with minimal price distortion, indicating greater liquidity, whereas a higher ratio suggests that even small trades cause significant price movements, signaling thinner trading conditions.

Figure \ref{fig:btc-marketcap} delineates three distinct groups of equities, differentiated by their BTC exposure $\beta$ (stock’s sensitivity to BTC returns), balance sheet intensity $\gamma$ (BTC holdings as a percentage of market capitalisation), and Amihud illiquidity ratio $\delta$ (trading liquidity):

\begin{enumerate}
    \item \textbf{High exposure $\beta$, medium intensity $\gamma$, high liquidity $|\log(\delta)|$ (upper region)}: Equities with $\beta > 1$ occupy the upper portion of the scatter plot. Their returns move more than one‑for‑one with BTC, displaying the strongest correlation. The extreme case is represented by MSTR, which appears as the highest point on the plot, reflecting the largest BTC‑equity correlation in the sample.
    \item \textbf{Low exposure $\beta$, low intensity $\gamma$, high liquidity $|\log(\delta)|$ (lower‑left region)}: Darker markers in the lower-left region represent equities whose returns have a low correlation with BTC. Their BTC holdings constitute only a small share of market capitalization and the equities are very liquid (low Amihud price impact). \textit{Tesla} (TSLA) exemplifies this group: although it holds $11,509$ BTC, the position is small relative to its market size (market cap of $\$1,025.5B$, but $\gamma = 0.1\%$), so the stock shows little co‑movement with the cryptocurrency.
    \item \textbf{Low exposure $\beta$, high intensity $\gamma$, low liquidity $|\log(\delta)|$ (lower-right region)}: A third set of equities lies toward the lower right part of the plot, indicating low correlation with BTC, even if they have high BTC/Market Cap. ratios. These stocks are comparatively illiquid (high Amihud impact) and therefore do not track BTC closely despite their sizeable balance sheet exposure. Limited trading depth dampens their measured correlation with BTC because even small orders can distort prices. An example is \textit{Fold Holdings Inc.} (FLD).
\end{enumerate}

\subsection{Rolling Transfer Entropy}\label{sec:rolling_transfer_entropy}

To further investigate the temporal dynamics and non-linear directional dependencies between BTC and equity markets, we extend our analysis by computing rolling window estimates of the TE between BTC and individual stock returns. Focusing on MSTR as an example of equity with significant Bitcoin exposure, the proposed high resolution approach enables an examination of how directional information flow evolves over time. 

\begin{figure}[h]
    \centering
    \includegraphics[width=\linewidth]{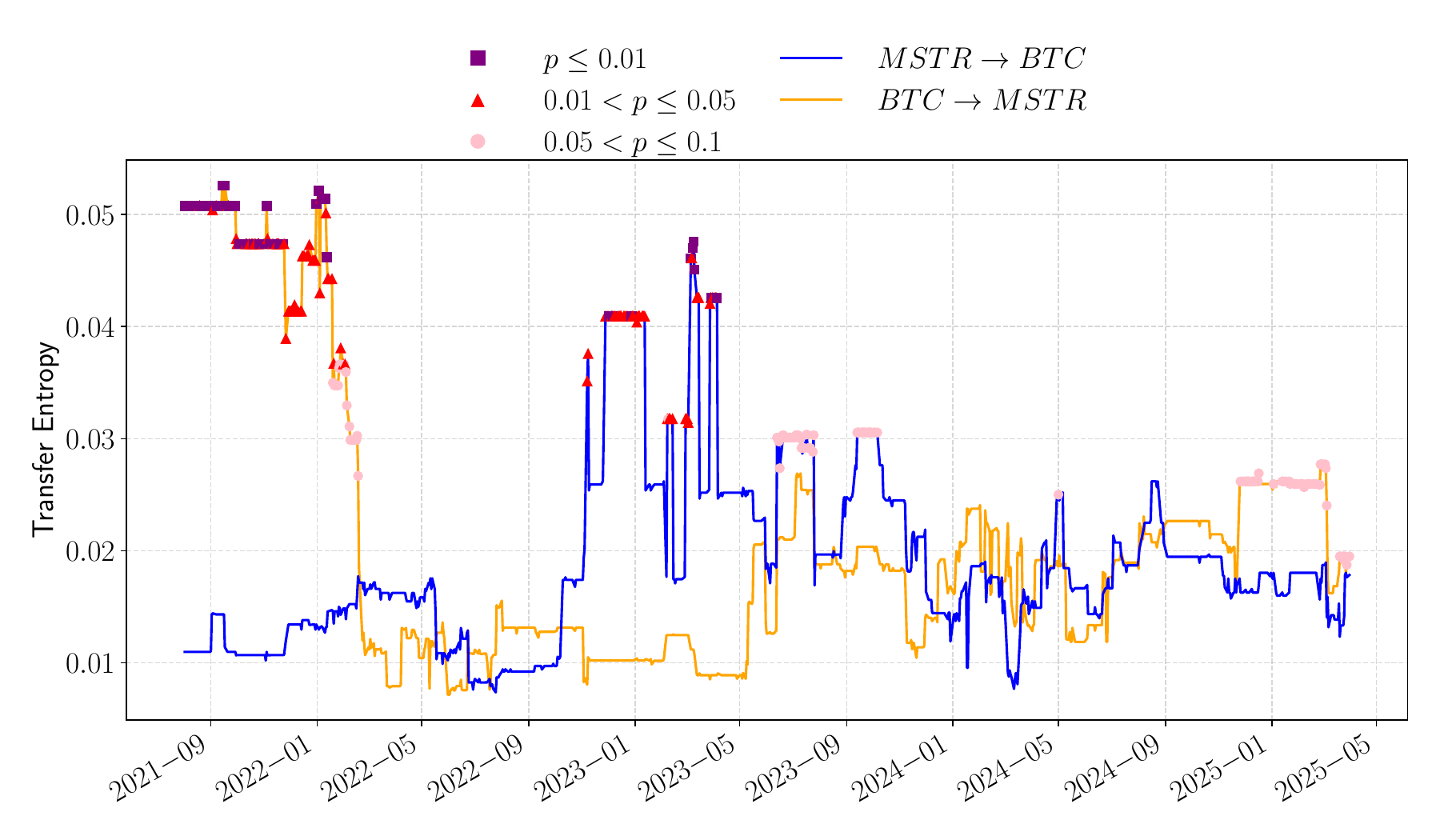}
    \caption{Rolling $252$ day TE between BTC and MSTR (August $2021$ to April $2025$). The chart plots one day lagged TE estimated in a $252$ trading day rolling window, obtained with $1000$ permutations. The blue line depicts information flow from MSTR returns to BTC returns (MSTR $\rightarrow$ BTC), and the orange line depicts information flow from BTC returns to MSTR returns (BTC $\rightarrow$ MSTR). Rolling estimates start in August $2021$, the earliest date for which a complete $252$ day window (extending back to MSTR’s inaugural BTC purchase in August $11$, $2020$) is available. Marker shape and colour convey statistical significance: purple squares for $p \le 0.01$, red triangles for $0.01 < p \le 0.05$, and pink circles for $0.05 < p \le 0.1$.}
    \label{fig:TE MSTR step size 1}
\end{figure}

From an initial inspection of Figure~\ref{fig:TE MSTR step size 1}, we observe considerable temporal variation and a pronounced asymmetry in the TE estimates. On average, TE$_{BTC \to MSTR}$ equals $0.0241$ bits, with a standard deviation of $0.0130$, whereas TE$_{MSTR \to BTC}$ is lower, averaging $0.0191$ bits with a standard deviation of $0.0074$. This asymmetry is further substantiated by statistical significance. Among the $1,298$ rolling windows analyzed, spanning from August $2021$ to April $2025$ with a daily stride, $410$ windows ($31.6\%$) exhibit statistically significant TE$_{BTC \to MSTR}$ at the $10\%$ significance level, compared to only $179$ windows ($13.8\%$) showing significant reverse TE.

These results align with the findings discussed in Section~\ref{sec:global measures}, where lagged correlations with equities leading BTC exhibited strong positive skewness ($1.78$) and excess kurtosis ($4.91$), indicative of rare but high impact episodes of predictive power. In contrast, lagged correlations with BTC leading equities displayed negative skewness ($-1.11$), suggesting a more consistent but moderate influence. The rolling TE analysis reinforces the view that directional dependencies between BTC and MSTR are dynamic and asymmetric, with BTC acting as a persistent source of informational influence, and reverse causality from MSTR to BTC appearing only sporadically and often linked to firm-specific events.

The TE$_{BTC \to MSTR}$ trend in Figure \ref{fig:TE MSTR step size 1} consistently exceeds the reverse direction and exhibits pronounced peaks during notable market events. These include the ETF speculation surge in late $2021$ \cite{coinbase2021futuresetf}, the official ETF approvals in January $2024$ \cite{yerushalmy2024bitcoinetf}, and the BTC halving event in April $2024$. In each case, the TE$_{BTC \to MSTR}$ exceeds $0.05$ bits with high statistical significance, indicating MSTR’s heightened sensitivity to BTC market signals. Conversely, TE$_{MSTR \to BTC}$ registers only intermittent peaks, notably between January and March $2023$ (coinciding with convertible bond issuance and BTC acquisitions \cite{miller2023microstrategy}), August to October $2023$ (following corporate finance announcements \cite{saylor2023interview}), and March to April $2025$ (during further BTC treasury purchases \cite{vanstraten2025strategy}). Even during these intervals, TE values rarely exceed $0.035$ bits, suggesting that MSTR’s influence on BTC remains limited in both magnitude and persistence and underlying the relatively minor role that firm-level actions play in shaping broader cryptocurrency market dynamics, especially when compared to systemic or macro-level developments driving BTC.

Outside these episodic bursts, both TE series converge near $0.02$ bits, indicating a baseline level of idiosyncratic noise and limited signal strength in the considered daily return data. The findings suggest that the directional dependency between BTC and MSTR is dynamic and asymmetric. BTC functions as a persistent informational driver, while reverse causality from MSTR to BTC is episodic and firm-specific.

Overall, the rolling TE analysis reinforces and enriches the interpretation offered by global metrics. While static averages suggest weak and approximately symmetric dependence, the rolling estimates unveil a temporally localized structure of information flow, bearing important implications for quantitative modeling and risk management.

\section{Conclusions}\label{sec:Conclusions}

This study offers a detailed empirical assessment of the evolving link between Bitcoin (BTC) and equity markets in the context of corporate treasury strategies centered on digital assets. Focusing on \textit{Strategy} (MSTR), the largest public holder of BTC, we combine correlation analysis, single factor models, and transfer entropy (TE) techniques to quantify the direction and dynamics of informational dependence between BTC and MSTR equity returns.

Our findings consistently indicate that BTC serves as the dominant source in the information flow. On average, TE$_{BTC \to MSTR}$ is higher than in the reverse direction, and statistically significant directional dependence from BTC to MSTR is more frequent and persistent. These asymmetries become particularly pronounced during market-wide events. In contrast, TE$_{MSTR \to BTC}$ peaks are rare, localized, and confined mainly to firm-specific actions such as convertible bond issuances or balance sheet disclosures.

These results suggest that MSTR operates primarily as a financial vehicle for BTC exposure, with limited influence in the opposite direction. Static dependency measures only partially capture the complex time-varying nature of these relationships. Rolling TE reveals a richer structure, characterized by intermittent bursts of significant influence and prolonged periods of near-random noise. This calls into question the reliability of static hedge ratios and highlights the need for adaptive quantitative strategies and risk management approaches that respond to changing market conditions.

Beyond the firm-level findings, our results point to a broader structural dynamic: the formation of feedback loops between corporate treasury strategies and financial market behavior. Firms like MSTR reflect BTC price movements in their equity valuations and reinforce BTC’s role in investor behavior and institutional positioning. However, this feedback remains highly asymmetric and unstable, with the direction of influence largely dictated by broader macro-financial developments rather than firm-level actions.

Several open challenges remain. Future research should focus on identifying leading indicators of episodes where equity-to-BTC feedback intensifies, which could support more responsive portfolio allocation, trading strategies, and policy oversight. Additional work could extend the analysis to a broader range of digital assets, assess intraday information flows using high-frequency data \cite{briola2024deep, briola2025hlob}, and incorporate entropy-based network models to evaluate multi-firm dynamics and systemic risk \cite{wang2023topological}. Moreover, emerging machine learning and artificial intelligence (AI) techniques \cite{wang2023homological, briola2023homological, wang2022network} offer promising tools for detecting nonlinear patterns, forecasting structural breaks, and adapting to shifting market regimes. Integrating these approaches with entropy-based measures could enhance financial interdependence models' real-time robustness.

As digital assets become more deeply embedded in corporate finance and capital markets, the capacity to monitor and interpret dynamic informational linkages will be essential for understanding asset pricing, managing systemic risk, and reassessing the boundaries of diversification in a rapidly evolving financial landscape.

\section{Acknowledgments}
All the authors acknowledge Hongyu Lin for the informal discussions that led to the genesis of the current research work. The authors S.A. and A.B. acknowledge Andrea Pezzella for the fruitful discussion on Section \ref{sec:sfm}. The author T.S. acknowledges Federico Peruzzo for the support in the parallelization of parts of the code. The author F.C. acknowledges support of the Economic and Social Research Council (ESRC) in funding the 17 Systemic Risk Centre at the LSE (ES/Y010612/1).

%\newpage
%\printbibliography
\bibliographystyle{plainnat}
\bibliography{reference}

\appendix
\section{Dataset}\label{app:dataset}
\small
% Please add the following required packages to your document preamble:
% \usepackage{booktabs}
% \usepackage[table,xcdraw]{xcolor}
% Beamer presentation requires \usepackage{colortbl} instead of \usepackage[table,xcdraw]{xcolor}
% \usepackage{longtable}
% Note: It may be necessary to compile the document several times to get a multi-page table to line up properly
\begin{longtable}[c]{cccccc}
\toprule
\rowcolor[HTML]{FFFFFF} 
{\color[HTML]{000000} \textbf{Name}} &
  {\color[HTML]{000000} \textbf{Ticker}} &
  {\color[HTML]{000000} \textbf{BTC}} &
  {\color[HTML]{000000} \textbf{Market Cap}} &
  {\color[HTML]{000000} \textbf{/M.Cap}} &
  {\color[HTML]{000000} \textbf{/21M}} \\* \midrule
\endfirsthead
\multicolumn{6}{c}%
{{\bfseries Table \thetable\ continued from previous page}} \\
\toprule
\rowcolor[HTML]{FFFFFF} 
{\color[HTML]{000000} \textbf{Name}} &
  {\color[HTML]{000000} \textbf{Ticker}} &
  {\color[HTML]{000000} \textbf{BTC}} &
  {\color[HTML]{000000} \textbf{Market Cap}} &
  {\color[HTML]{000000} \textbf{/M.Cap}} &
  {\color[HTML]{000000} \textbf{/21M}} \\* \midrule
\endhead
\rowcolor[HTML]{EFEFEF} 
{\color[HTML]{000000} Microstrategy, Inc.} &
  {\color[HTML]{000000} MSTR} &
  {\color[HTML]{000000} 528,185} &
  {\color[HTML]{000000} \$76,126 M} &
  {\color[HTML]{000000} 56.97\%} &
  {\color[HTML]{000000} 2.515\%} \\* \midrule
\rowcolor[HTML]{EFEFEF} 
{\color[HTML]{000000} MARA Holdings, Inc.} &
  {\color[HTML]{000000} MARA} &
  {\color[HTML]{000000} 47,600} &
  {\color[HTML]{000000} \$4,169 M} &
  {\color[HTML]{000000} 93.76\%} &
  {\color[HTML]{000000} 0.227\%} \\* \midrule
\rowcolor[HTML]{EFEFEF} 
{\color[HTML]{000000} Riot Platforms, Inc.} &
  {\color[HTML]{000000} RIOT} &
  {\color[HTML]{000000} 19,223} &
  {\color[HTML]{000000} \$2,432 M} &
  {\color[HTML]{000000} 64.99\%} &
  {\color[HTML]{000000} 0.092\%} \\* \midrule
\rowcolor[HTML]{EFEFEF} 
{\color[HTML]{000000} CleanSpark, Inc.} &
  {\color[HTML]{000000} CLSK} &
  {\color[HTML]{000000} 11,869} &
  {\color[HTML]{000000} \$2,051 M} &
  {\color[HTML]{000000} 47.51\%} &
  {\color[HTML]{000000} 0.057\%} \\* \midrule
\rowcolor[HTML]{EFEFEF} 
{\color[HTML]{000000} Tesla, Inc.} &
  {\color[HTML]{000000} TSLA} &
  {\color[HTML]{000000} 11,509} &
  {\color[HTML]{000000} \$783,995 M} &
  {\color[HTML]{000000} 0.12\%} &
  {\color[HTML]{000000} 0.055\%} \\* \midrule
\rowcolor[HTML]{EFEFEF} 
{\color[HTML]{000000} Hut 8 Mining Corp} &
  {\color[HTML]{000000} HUT} &
  {\color[HTML]{000000} 10,273} &
  {\color[HTML]{000000} \$1,226 M} &
  {\color[HTML]{000000} 68.83\%} &
  {\color[HTML]{000000} 0.049\%} \\* \midrule
\rowcolor[HTML]{EFEFEF} 
{\color[HTML]{000000} Coinbase Global, Inc.} &
  {\color[HTML]{000000} COIN} &
  {\color[HTML]{000000} 9,480} &
  {\color[HTML]{000000} \$42,811 M} &
  {\color[HTML]{000000} 1.82\%} &
  {\color[HTML]{000000} 0.045\%} \\* \midrule
\rowcolor[HTML]{EFEFEF} 
{\color[HTML]{000000} Block, Inc.} &
  {\color[HTML]{000000} XYZ} &
  {\color[HTML]{000000} 8,485} &
  {\color[HTML]{000000} \$31,950 M} &
  {\color[HTML]{000000} 2.18\%} &
  {\color[HTML]{000000} 0.040\%} \\* \midrule
\rowcolor[HTML]{EFEFEF} 
{\color[HTML]{000000} Metaplanet Inc.} &
  {\color[HTML]{000000} 3350.T} &
  {\color[HTML]{000000} 4,206} &
  {\color[HTML]{000000} \$981 M} &
  {\color[HTML]{000000} 35.21\%} &
  {\color[HTML]{000000} 0.020\%} \\* \midrule
%\rowcolor[HTML]{EFEFEF} 
%{\color[HTML]{000000} Bitcoin Group SE} &
  %{\color[HTML]{000000} ADE.DE} &
  %{\color[HTML]{000000} 3,605} &
  %{\color[HTML]{000000} \$179 M} &
  %{\color[HTML]{000000} 165.36\%} &
  %{\color[HTML]{000000} 0.017\%} \\* \midrule
\rowcolor[HTML]{EFEFEF} 
{\color[HTML]{000000} Semler Scientific} &
  {\color[HTML]{000000} SMLR} &
  {\color[HTML]{000000} 3,192} &
  {\color[HTML]{000000} \$304 M} &
  {\color[HTML]{000000} 86.21\%} &
  {\color[HTML]{000000} 0.015\%} \\* \midrule
\rowcolor[HTML]{EFEFEF} 
{\color[HTML]{000000} Boyaa Interactive International Limited} &
  {\color[HTML]{000000} 0434.HK} &
  {\color[HTML]{000000} 3,183} &
  {\color[HTML]{000000} \$330 M} &
  {\color[HTML]{000000} 79.13\%} &
  {\color[HTML]{000000} 0.015\%} \\* \midrule
\rowcolor[HTML]{EFEFEF} 
{\color[HTML]{000000} Galaxy Digital Holdings Ltd} &
  {\color[HTML]{000000} GLXY. TO} &
  {\color[HTML]{000000} 3,150} &
  {\color[HTML]{000000} \$3,594 M} &
  {\color[HTML]{000000} 7.20\%} &
  {\color[HTML]{000000} 0.015\%} \\* \midrule
\rowcolor[HTML]{EFEFEF} 
{\color[HTML]{000000} HIVE Digital Technologies} &
  {\color[HTML]{000000} HIVE} &
  {\color[HTML]{000000} 2,620} &
  {\color[HTML]{000000} \$225 M} &
  {\color[HTML]{000000} 95.48\%} &
  {\color[HTML]{000000} 0.012\%} \\* \midrule
\rowcolor[HTML]{EFEFEF} 
{\color[HTML]{000000} Cango Inc} &
  {\color[HTML]{000000} CANG} &
  {\color[HTML]{000000} 2,475} &
  {\color[HTML]{000000} \$363 M} &
  {\color[HTML]{000000} 55.96\%} &
  {\color[HTML]{000000} 0.012\%} \\* \midrule
\rowcolor[HTML]{EFEFEF} 
{\color[HTML]{000000} Exodus Movement, Inc} &
  {\color[HTML]{000000} EXOD} &
  {\color[HTML]{000000} 1,900} &
  {\color[HTML]{000000} \$1,150 M} &
  {\color[HTML]{000000} 13.57\%} &
  {\color[HTML]{000000} 0.009\%} \\* \midrule
\rowcolor[HTML]{EFEFEF} 
{\color[HTML]{000000} BITFUFU} &
  {\color[HTML]{000000} FUFU} &
  {\color[HTML]{000000} 1,800} &
  {\color[HTML]{000000} \$699 M} &
  {\color[HTML]{000000} 21.16\%} &
  {\color[HTML]{000000} 0.009\%} \\* \midrule
\rowcolor[HTML]{EFEFEF} 
{\color[HTML]{000000} NEXON Co., Ltd.} &
  {\color[HTML]{000000} 3659.T} &
  {\color[HTML]{000000} 1,717} &
  {\color[HTML]{000000} \$11,831 M} &
  {\color[HTML]{000000} 1.19\%} &
  {\color[HTML]{000000} 0.008\%} \\* \midrule
\rowcolor[HTML]{EFEFEF} 
{\color[HTML]{000000} Fold Holdings Inc.} &
  {\color[HTML]{000000} FLD} &
  {\color[HTML]{000000} 1,485} &
  {\color[HTML]{000000} \$126 M} &
  {\color[HTML]{000000} 96.98\%} &
  {\color[HTML]{000000} 0.007\%} \\* \midrule
\rowcolor[HTML]{EFEFEF} 
{\color[HTML]{000000} Cipher Mining} &
  {\color[HTML]{000000} CIFR} &
  {\color[HTML]{000000} 1,344} &
  {\color[HTML]{000000} \$826 M} &
  {\color[HTML]{000000} 13.37\%} &
  {\color[HTML]{000000} 0.006\%} \\* \midrule
\rowcolor[HTML]{EFEFEF} 
{\color[HTML]{000000} Canaan Inc.} &
  {\color[HTML]{000000} CAN} &
  {\color[HTML]{000000} 1,231} &
  {\color[HTML]{000000} \$243 M} &
  {\color[HTML]{000000} 41.55\%} &
  {\color[HTML]{000000} 0.006\%} \\* \midrule
\rowcolor[HTML]{EFEFEF} 
{\color[HTML]{000000} Aker ASA} &
  {\color[HTML]{000000} AKER.OL} &
  {\color[HTML]{000000} 1,170} &
  {\color[HTML]{000000} \$3,885 M} &
  {\color[HTML]{000000} 2.47\%} &
  {\color[HTML]{000000} 0.006\%} \\* \midrule
\rowcolor[HTML]{EFEFEF} 
{\color[HTML]{000000} Bitfarms Ltd.} &
  {\color[HTML]{000000} BITF} &
  {\color[HTML]{000000} 1,152} &
  {\color[HTML]{000000} \$437 M} &
  {\color[HTML]{000000} 21.67\%} &
  {\color[HTML]{000000} 0.005\%} \\* \midrule
\rowcolor[HTML]{EFEFEF} 
{\color[HTML]{000000} Bitdeer Technologies Group} &
  {\color[HTML]{000000} BTDR} &
  {\color[HTML]{000000} 1,143} &
  {\color[HTML]{000000} \$1,471 M} &
  {\color[HTML]{000000} 6.38\%} &
  {\color[HTML]{000000} 0.005\%} \\* \midrule
\rowcolor[HTML]{EFEFEF} 
{\color[HTML]{000000} Ming Shing Group} &
  {\color[HTML]{000000} MSW} &
  {\color[HTML]{000000} 833} &
  {\color[HTML]{000000} \$45 M} &
  {\color[HTML]{000000} 152.80\%} &
  {\color[HTML]{000000} 0.004\%} \\* \midrule
\rowcolor[HTML]{EFEFEF} 
{\color[HTML]{000000} Bit Digital, Inc.} &
  {\color[HTML]{000000} BTBT} &
  {\color[HTML]{000000} 742.1} &
  {\color[HTML]{000000} \$325 M} &
  {\color[HTML]{000000} 18.73\%} &
  {\color[HTML]{000000} 0.004\%} \\* \midrule
BlackRock, Inc. &
  BLK &
  6.15 &
  \$134,558 M &
  0.00\% &
  0.000\% \\* \midrule
MercadoLibre, Inc. &
  MELI &
  412.7 &
  \$99,707 M &
  0.03\% &
  0.002\% \\* \midrule
Intesa Sanpaolo &
  ISP.MI &
  11 &
  \$83,004 M &
  0.00\% &
  0.000\% \\* \midrule
Globant S.A. &
  GLOB &
  15 &
  \$4,624 M &
  0.03\% &
  0.000\% \\* \midrule
Rumble Inc. &
  RUM &
  188 &
  \$2,512 M &
  0.61\% &
  0.001\% \\* \midrule
Core Scientific &
  CORZ &
  21.02 &
  \$2,060 M &
  0.08\% &
  0.000\% \\* \midrule
\rowcolor[HTML]{EFEFEF} 
LM Funding America &
  LMFA &
  160.5 &
  \$6 M &
  230.11\% &
  0.001\% \\* \midrule
\rowcolor[HTML]{EFEFEF} 
Advanced Bitcoin Technologies AG &
  ABT.DU &
  242.2 &
  \$9 M &
  210.75\% &
  0.001\% \\* \midrule
\rowcolor[HTML]{EFEFEF} 
Genius Group &
  GNS &
  440 &
  \$19 M &
  191.61\% &
  0.002\% \\* \midrule
\rowcolor[HTML]{EFEFEF} 
DMG Blockchain Solutions Inc. &
  DMGI.V &
  423 &
  \$30 M &
  116.77\% &
  0.002\% \\* \midrule
\rowcolor[HTML]{EFEFEF} 
The Blockchain Group &
  ALTBG.PA &
  620 &
  \$52 M &
  97.97\% &
  0.003\% \\* \midrule
\rowcolor[HTML]{EFEFEF} 
LQWD Technologies Corp. &
  LQWD.V &
  161 &
  \$18 M &
  74.15\% &
  0.001\% \\* \midrule
\rowcolor[HTML]{EFEFEF} 
Cathedra Bitcoin Inc. &
  CBIT.V &
  52.5 &
  \$6 M &
  70.55\% &
  0.000\% \\* \midrule
\rowcolor[HTML]{EFEFEF} 
SATO Technologies Corp &
  SATO.V &
  36 &
  \$6 M &
  51.49\% &
  0.000\% \\* \bottomrule
\caption{Dataset of $39$ BTC holding entities updated at April 11, 2025. The Table presents the dataset used in our analysis: the first group (highlighted in grey) is ordered by decreasing BTC holdings, down to 700 BTC -- \textit{BTC} column. The second group (unshaded rows) is ordered by decreasing Market Capitalization, down to a threshold of $\$1,000$ million -- \textit{Market Cap} column. The final group is ordered by decreasing BTC holdings as a percentage of market capitalization (/M.Cap), down to entities for which this ratio is equal to $50.00\%$ -- \textit{/M.Cap} column. The last column -- \textit{/21M} --  represents BTC holdings as a percentage of total BTC supply.}
\label{tab:tab dataset}\\
\end{longtable}

\section{Stocks' Daily Log Returns}\label{app:daily returns}
\small
% Please add the following required packages to your document preamble:
% \usepackage{booktabs}
% \usepackage{longtable}
% Note: It may be necessary to compile the document several times to get a multi-page table to line up properly
\begin{longtable}[c]{@{}cccccccc@{}}
\toprule
\textbf{Ticker} & \textbf{Mean} & \textbf{Median} & \textbf{Std} & \textbf{Skewness} & \textbf{Kurtosis} & \textbf{Min.} & \textbf{Max.} \\* \midrule
\endfirsthead
\multicolumn{8}{c}%
{{\bfseries Table \thetable\ continued from previous page}} \\
\toprule
\textbf{Ticker} & \textbf{Mean} & \textbf{Median} & \textbf{Std} & \textbf{Skewness} & \textbf{Kurtosis} & \textbf{Min.} & \textbf{Max.} \\* \midrule
\endhead
BTC-USD	& 0.0015 &	0.0002	& 0.0252	& 0.3056	& 2.3004	& -0.0908 &	0.1146  \\* \midrule
MSTR     & 0.0047  & 0.0007  & 0.0582 & 0.1576  & 1.5407  & -0.2384 & 0.229  \\* \midrule
MARA     & 0.0007  & -0.003  & 0.065  & 0.3773  & 1.1615  & -0.1807 & 0.2618 \\* \midrule
RIOT     & -0.0005 & -0.0038 & 0.0579 & 0.3206  & 0.9352  & -0.1722 & 0.2324 \\* \midrule
CLSK     & 0.0021  & -0.0065 & 0.0681 & 0.6031  & 1.1119  & -0.1798 & 0.2841 \\* \midrule
TSLA     & 0.0006  & 0.0003  & 0.0371 & 0.2055  & 3.2978  & -0.1675 & 0.1982 \\* \midrule
HUT      & 0.0008  & -0.0022 & 0.0646 & 0.0359  & 1.3314  & -0.275  & 0.2276 \\* \midrule
COIN     & 0.002   & -0.0021 & 0.0501 & 0.5214  & 2.712   & -0.1933 & 0.2709 \\* \midrule
XYZ      & -0.0004 & 0.0009  & 0.0307 & -0.4856 & 5.1462  & -0.1947 & 0.1495 \\* \midrule
3350.T   & 0.0053  & 0.0     & 0.0872 & 1.0756  & 8.2263  & -0.3102 & 0.6391 \\* \midrule
SMLR     & 0.001   & -0.0003 & 0.0503 & 0.3437  & 7.4893  & -0.2916 & 0.2691 \\* \midrule
0434.HK  & 0.0042  & 0.0     & 0.0542 & 1.5023  & 7.5087  & -0.1811 & 0.3577 \\* \midrule
GLXY.TO  & 0.0022  & 0.0     & 0.0488 & 0.1912  & 1.6364  & -0.1834 & 0.2296 \\* \midrule
HIVE     & -0.0016 & -0.0033 & 0.0537 & 0.1708  & 0.8752  & -0.2291 & 0.1814 \\* \midrule
CANG     & 0.0024  & 0.0     & 0.0475 & 0.1417  & 7.9527  & -0.3285 & 0.23   \\* \midrule
EXOD     & 0.0057  & 0.0     & 0.1384 & 1.3495  & 15.4365 & -0.7658 & 1.0282 \\* \midrule
FUFU     & -0.0016 & 0.0     & 0.0707 & 2.2127  & 29.5002 & -0.419  & 0.7217 \\* \midrule
3659.T   & -0.0009 & 0.0001  & 0.0282 & -0.6789 & 13.985  & -0.1918 & 0.1962 \\* \midrule
FLD      & -0.0013 & 0.0     & 0.0288 & 0.4125  & 51.227  & -0.2834 & 0.2712 \\* \midrule
CIFR     & 0.0     & -0.0062 & 0.0712 & 0.3772  & 1.2751  & -0.2687 & 0.2701 \\* \midrule
CAN      & -0.0022 & -0.006  & 0.07   & 0.6128  & 3.5542  & -0.3112 & 0.3455 \\* \midrule
AKER.OL  & -0.0001 & 0.0     & 0.014  & 0.1105  & 1.5164  & -0.0522 & 0.056  \\* \midrule
BITF     & -0.0003 & -0.0085 & 0.0587 & 0.5352  & 0.9244  & -0.172  & 0.2344 \\* \midrule
BTDR     & -0.0004 & -0.0014 & 0.0834 & 0.1194  & 3.1993  & -0.3524 & 0.4116 \\* \midrule
MSW      & -0.0043 & -0.0015 & 0.1154 & -1.5949 & 9.5066  & -0.6231 & 0.3298 \\* \midrule
BTBT     & 0.0007  & -0.0037 & 0.0652 & 0.4878  & 1.6968  & -0.2057 & 0.3344 \\* \midrule
BLK      & 0.0008  & 0.0011  & 0.0129 & 0.0443  & 1.7629  & -0.0591 & 0.0529 \\* \midrule
MELI     & 0.0008  & 0.0011  & 0.0243 & -0.5459 & 7.8564  & -0.1769 & 0.1274 \\* \midrule
ISP.MI   & 0.0018  & 0.0019  & 0.0134 & -0.8582 & 4.6997  & -0.0907 & 0.0362 \\* \midrule
GLOB     & -0.0006 & 0.0002  & 0.0284 & -2.9641 & 35.2498 & -0.3259 & 0.112  \\* \midrule
RUM      & -0.0004 & -0.004  & 0.0564 & 3.2843  & 28.1806 & -0.151  & 0.5946 \\* \midrule
CORZ     & 0.0028  & 0.0017  & 0.0599 & -0.1892 & 6.5011  & -0.3483 & 0.3382 \\* \midrule
LMFA     & -0.0026 & -0.0024 & 0.0695 & 0.4687  & 2.9778  & -0.2624 & 0.3363 \\* \midrule
ABT.DU   & -0.0005 & 0.0     & 0.1145 & 0.0591  & 5.5089  & -0.4601 & 0.7057 \\* \midrule
GNS      & -0.0083 & -0.0143 & 0.0957 & 0.2967  & 9.29    & -0.6796 & 0.5092 \\* \midrule
DMGI.V   & -0.0006 & 0.0     & 0.0587 & 0.7454  & 2.9376  & -0.1728 & 0.3327 \\* \midrule
ALTBG.PA & 0.0013  & 0.0     & 0.0591 & 0.3647  & 8.6981  & -0.3448 & 0.3192 \\* \midrule
LQWD.V   & 0.0007  & 0.0     & 0.078  & 0.4967  & 3.4193  & -0.3001 & 0.3959 \\* \midrule
CBIT.V   & -0.0058 & 0.0     & 0.081  & 0.2099  & 2.1244  & -0.2877 & 0.3137 \\* \midrule
SATO.V   & -0.001  & 0.0     & 0.0599 & 0.7779  & 6.1668  & -0.2311 & 0.3395 \\* \bottomrule
\caption{Descriptive statistics of daily log returns, from April $1$, $2023$ to April $1$, $2025$.}
\label{tab:stats}\\
\end{longtable}

\section{Global Measures of Information}\label{app:correlation}

% Please add the following required packages to your document preamble:
% \usepackage{booktabs}
% \usepackage{longtable}
% Note: It may be necessary to compile the document several times to get a multi-page table to line up properly
{\scriptsize
\begin{longtable}[c]{@{}cccccc@{}}
\toprule
\textbf{Ticker} &
  \textbf{\begin{tabular}[c]{@{}c@{}}PC: \\ BTC (t) $\rightleftarrows$ X(t)\end{tabular}} &
  \textbf{\begin{tabular}[c]{@{}c@{}}PC: \\ BTC (t) $\rightleftarrows$ X(t-1)\end{tabular}} &
  \textbf{\begin{tabular}[c]{@{}c@{}}PC: \\ BTC (t-1) $\rightleftarrows$ X(t)\end{tabular}} &
  \textbf{\begin{tabular}[c]{@{}c@{}}TE: \\ BTC(t-1) $\rightarrow$ X(t)\end{tabular}} &
  \textbf{\begin{tabular}[c]{@{}c@{}}TE: \\ X(t-1) $\rightarrow$ BTC(t)\end{tabular}} \\* \midrule
\endfirsthead
\multicolumn{6}{c}%
{{\bfseries Table \thetable\ continued from previous page}} \\
\toprule
\textbf{Ticker} &
  \textbf{\begin{tabular}[c]{@{}c@{}}PC: \\ BTC (t) $\rightleftarrows$ X(t)\end{tabular}} &
  \textbf{\begin{tabular}[c]{@{}c@{}}PC: \\ BTC (t) $\rightleftarrows$ X(t-1)\end{tabular}} &
  \textbf{\begin{tabular}[c]{@{}c@{}}PC: \\ BTC (t-1) $\rightleftarrows$ X(t)\end{tabular}} &
  \textbf{\begin{tabular}[c]{@{}c@{}}TE: \\ BTC(t-1) $\rightarrow$ X(t)\end{tabular}} &
  \textbf{\begin{tabular}[c]{@{}c@{}}TE: \\ X(t-1) $\rightarrow$ BTC(t)\end{tabular}} \\* \midrule
\endhead
MSTR     & 0.660** & -0.031   & -0.011   & 0.020*  & 0.006   \\* \midrule
HIVE     & 0.580** & 0.007    & -0.007   & 0.012   & 0.023*  \\* \midrule
MARA     & 0.579** & 0.014    & 0.002    & 0.010   & 0.013   \\* \midrule
RIOT     & 0.577** & 0.012    & -0.014   & 0.006   & 0.013   \\* \midrule
COIN     & 0.572** & -0.048   & 0.004    & 0.009   & 0.013   \\* \midrule
HUT      & 0.556** & -0.031   & -0.033   & 0.011   & 0.019   \\* \midrule
CLSK     & 0.545** & -0.021   & 0.033    & 0.018*  & 0.021*  \\* \midrule
BTBT     & 0.539** & 0.009    & -0.008   & 0.017   & 0.016   \\* \midrule
GLXY.TO  & 0.533** & 0        & 0.014    & 0.002   & 0.014   \\* \midrule
BITF     & 0.527** & -0.009   & -0.033   & 0.015   & 0.017   \\* \midrule
DMGI.V   & 0.499** & -0.049   & -0.090** & 0.011   & 0.011   \\* \midrule
CAN      & 0.443** & -0.133** & 0.004    & 0.013   & 0.022*  \\* \midrule
CIFR     & 0.442** & 0.005    & 0.026    & 0.014   & 0.019*  \\* \midrule
CORZ     & 0.422** & 0.013    & -0.068   & 0.020   & 0.014   \\* \midrule
LMFA     & 0.391** & -0.056   & 0.037    & 0.025*  & 0.013   \\* \midrule
CBIT.V   & 0.324** & -0.041   & -0.147*  & 0.018   & 0.011   \\* \midrule
BTDR     & 0.314** & 0.022    & -0.024   & 0.009   & 0.016   \\* \midrule
SMLR     & 0.297** & 0.005    & 0.032    & 0.018   & 0.013   \\* \midrule
XYZ      & 0.290** & 0.004    & 0.085*   & 0.007   & 0.014   \\* \midrule
LQWD.V   & 0.254** & -0.025   & 0.042    & 0.018   & 0.010   \\* \midrule
TSLA     & 0.250** & -0.107** & 0.077*   & 0.008   & 0.016   \\* \midrule
FUFU     & 0.245** & -0.096** & -0.049   & 0.018   & 0.018   \\* \midrule
BLK      & 0.203** & -0.02    & 0.066    & 0.008   & 0.014   \\* \midrule
SATO.V   & 0.181** & -0.034   & 0.031    & 0.023*  & 0.010   \\* \midrule
RUM      & 0.171** & 0.029    & 0.016    & 0.022** & 0.015   \\* \midrule
ALTBG.PA & 0.144** & -0.009   & 0.03     & 0.014   & 0.006   \\* \midrule
GLOB     & 0.134** & -0.006   & -0.01    & 0.001   & 0.012   \\* \midrule
0434.HK  & 0.133** & -0.014   & 0.283**  & 0.032** & 0.027** \\* \midrule
GNS      & 0.107** & -0.022   & 0.080*   & 0.028** & 0.015   \\* \midrule
EXOD     & 0.102** & -0.100** & -0.026   & 0.013   & 0.003   \\* \midrule
CANG     & 0.100** & -0.014   & 0.065    & 0.022   & 0.004   \\* \midrule
MELI     & 0.088** & -0.042   & -0.006   & 0.016   & 0.011   \\* \midrule
FLD      & 0.076*  & 0.037    & -0.008   & 0.014   & 0.010   \\* \midrule
3659.T   & 0.053   & -0.017   & -0.018   & 0.014   & 0.010   \\* \midrule
ISP.MI   & 0.042   & 0.035    & -0.052   & 0.014   & 0.016   \\* \midrule
3350.T   & 0.038   & 0.035    & 0.283**  & 0.020   & 0.008   \\* \midrule
ABT.DU   & 0.025   & 0.011    & -0.001   & 0.021   & 0.012   \\* \midrule
MSW      & -0.026  & 0.023    & 0.042    & 0.016   & 0.033   \\* \midrule
AKER.OL  & -0.035  & -0.031   & 0.051    & 0.013   & 0.023** \\* \bottomrule
\caption{Measures of information between BTC and individual stocks. The Table reports multiple measures of dependence between BTC and $39$ stocks ($X$), computed over the period from April $1$, $2023$ to April $1$, $2025$. The columns two to four present Pearson Correlation (PC) coefficients: (i) between BTC and each stock at the same time $t$; (ii) between BTC at time $t$ and the stock at time $t-1$; and (iii) between BTC at time $t-1$ and the stock at time $t$. The final two columns report Transfer Entropy (TE) estimates with a one day lag, capturing potential directional information flow from BTC to the stock $X$ and vice versa. Statistical significance is indicated with stars: * $0.05 < p <0.1$; ** $p \leq 0.05$. The Table is sorted in decreasing order based on the contemporaneous Pearson Correlation between BTC and each stock.}
\label{tab:table_correlations}\\
\end{longtable}}

\section{Single Factor Model}\label{app:sfm}
\small
% Please add the following required packages to your document preamble:
% \usepackage{booktabs}
% \usepackage{longtable}
% Note: It may be necessary to compile the document several times to get a multi-page table to line up properly
\begin{longtable}[c]{@{}cccc@{}}
\toprule
\textbf{Ticker} & \textbf{$\beta$} & \textbf{$\alpha$} & \textbf{$R^2$} \\* \midrule
\endfirsthead
\endhead
MSTR             & 1.36             & 0.002             & 0.44           \\* \midrule
MARA             & 1.34             & -0.0019           & 0.34           \\* \midrule
CLSK             & 1.32             & -0.00051          & 0.3            \\* \midrule
HUT              & 1.27             & -0.0017           & 0.31           \\* \midrule
BTBT             & 1.24             & -0.0018           & 0.29           \\* \midrule
RIOT             & 1.19             & -0.0028           & 0.33           \\* \midrule
CIFR             & 1.12             & -0.0021           & 0.2            \\* \midrule
HIVE             & 1.11             & -0.0037           & 0.34           \\* \midrule
CAN              & 1.10             & -0.0044           & 0.2            \\* \midrule
BITF             & 1.10             & -0.0024           & 0.28           \\* \midrule
DMGI.V           & 1.06             & -0.0025           & 0.25           \\* \midrule
COIN             & 1.02             & 3.4e-05           & 0.33           \\* \midrule
LMFA             & 0.96             & -0.0045           & 0.15           \\* \midrule
GLXY.TO          & 0.94             & 0.00053           & 0.28           \\* \midrule
BTDR             & 0.93             & -0.0022           & 0.098          \\* \midrule
CBIT.V           & 0.9              & -0.0083           & 0.11           \\* \midrule
CORZ             & 0.83             & 0.00086           & 0.18           \\* \midrule
LQWD.V           & 0.72             & -0.00061          & 0.065          \\* \midrule
FUFU             & 0.62             & -0.0028           & 0.06           \\* \midrule
SMLR             & 0.53             & -6.6e-05          & 0.088          \\* \midrule
EXOD            & 0.50              & 4.7e-03           & 0.010         \\* \midrule
SATO.V           & 0.39             & -0.0017           & 0.033          \\* \midrule
GNS              & 0.36             & -9.0e-03          & 0.012         \\* \midrule
RUM              & 0.34             & -0.0011           & 0.029          \\* \midrule
TSLA             & 0.33             & 7e-07             & 0.062          \\* \midrule
XYZ              & 0.32             & -0.001            & 0.084          \\* \midrule
ALTBG.PA         & 0.31             & 0.00063           & 0.021          \\* \midrule
0434.HK          & 0.26             & 0.0038            & 0.018          \\* \midrule
CANG             & 0.17             & 0.002             & 0.01           \\* \midrule
GLOB             & 0.14             & -0.00089          & 0.018          \\* \midrule
3350.T           & 0.12             & 0.0051            & 0.0015         \\* \midrule
ABT.DU          & 0.10              & -7.5e-04          & 0.00065       \\* \midrule
BLK              & 0.093            & 0.00062           & 0.041          \\* \midrule
FLD              & 0.077            & -0.0014           & 0.0057         \\* \midrule
MELI             & 0.076            & 0.00061           & 0.0077         \\* \midrule
3659.T           & 0.054            & -0.00095          & 0.0028         \\* \midrule
ISP.MI           & 0.02             & 0.0017            & 0.0017         \\* \midrule
AKER.OL          & -0.018           & -8.3e-05          & 0.0012         \\* \midrule
MSW             & -0.11             & -4.39e-03         & 0.00067       \\* \bottomrule
\caption{Single Factor regression results (1 Apr 2023 - 1 Apr 2025) for the $39$ companies of our dataset, sorted with decreasing $\beta$. For each company, the table reports the estimated BTC \(\beta\), \(\alpha\), and coefficient \(R^{2}\) from the regression in Equation \ref{eq:regression}.  Betas quantify the sensitivity of each equity to daily BTC moves, whereas \(R^{2}\) indicates the share of return variance explained by the single factor model.}
\label{tab:tab betas}\\
\end{longtable}

\end{document}